\documentclass[3p,times]{elsarticle}

\usepackage{amsfonts}
\usepackage{mathrsfs}
\usepackage{amsmath}
\usepackage{amssymb}

 \usepackage{lineno}
\newtheorem{theorem}{\indent Theorem}[section]

\newtheorem{lemma}{\indent Lemma}[section]
\newtheorem{remark}{\indent Remark}[section]

\journal{Discrete and Continuous Dynamical Systems}

\begin{document}

\begin{frontmatter}



\title{Global and exponential attractors of the three dimensional viscous primitive equations of large-scale moist atmosphere}


\author{Bo You$^{a,*}$, Fang Li$^b$}

\address{$^a$ School of Mathematics and Statistics, Xi'an Jiaotong University, Xi'an, 710049, P. R. China\\
$^b$ School of Mathematics and Statistics, Xidian University, Xi'an, 710126, P. R. China}

 \cortext[cor1]{Corresponding author.\\
 E-mail addresses:youb2013@xjtu.edu.cn(B. You),fli@xidian.edu.cn (F. Li)}

\begin{abstract}
This paper is concerned with the long-time behavior of solutions for the three dimensional viscous
primitive equations of large-scale moist atmosphere. We prove the existence of a global attractor in $(H^2(\Omega))^4\cap V$ for the three dimensional viscous primitive equations of large-scale moist atmosphere by asymptotic a priori estimate and construct an exponential attractor by using the smoothing property of the semigroup generated by problem \eqref{2.4}-\eqref{2.12}. As a byproduct, we obtain the fractal dimension of the global attractor for the semigroup generated by problem \eqref{2.4}-\eqref{2.12} is finite, which is in consistent with the results in \cite{jn2, jn1}.
\end{abstract}


\end{frontmatter}


\section{Introduction}
\def\theequation{1.\arabic{equation}}\makeatother
\setcounter{equation}{0}
In this paper, we consider the long-time behavior of solutions for the following three dimensional viscous primitive equations of large-scale moist atmosphere in the pressure
coordinate system(see \cite{gbl2, gbl3, pj1, vgk})
\begin{align}\label{1.1}
&\frac{\partial v}{\partial t}+(v\cdot \nabla)v+w\frac{\partial
v}{\partial z}+\nabla
\Phi+\frac{1}{Ro}fv^{\bot}+L_1v=0,\\
\label{1.2}&\frac{\partial \Phi}{\partial z}+\frac{bP}{p}(1+aq)T=0,\\
\label{1.3}&\nabla\cdot v+\frac{\partial w}{\partial z}=0,\\
\label{1.4}&\frac{\partial T}{\partial t}+v\cdot \nabla
T+w\frac{\partial T}{\partial z}-\frac{bP}{p}(1+aq)w+L_2T=Q_1,\\
\label{1.5}&\frac{\partial q}{\partial t}+v\cdot \nabla
q+w\frac{\partial q}{\partial z}+L_3q=Q_2
\end{align}
in the domain
\begin{align*}
\Omega=M\times(0,1),
\end{align*}
where $M$ is a bounded domain in $\mathbb{R}^2$ with smooth boundary $\partial M$. The unknown functions for problem \eqref{1.1}-\eqref{1.5} are
the horizontal velocity field $v=(v_1,v_2)$, the vertical
velocity $w$ in $p$-coordinate system, the mixing ratio of water vapor in the air $q$, the
temperature $T$ and the geopotential $\Phi.$  Here $v^{\bot}=(-v_2,v_1),$  $f=2\cos\theta_0$ is the Coriolis parameter,
$Ro$ is the Rossby number, $P$ is an approximate value of pressure at the surface of the
earth, $p_0$ represents the pressure of the upper atmosphere and $p_0 > 0,$ the variable $z$ satisfies $p =(P- p_0)z+ p_0$
$(0 < p_0\leq p\leq P),$ $Q_1,$ $Q_2$ are given functions on $\Omega$ (here we don't consider the condensation
of water vapor), $a,$ $b$ are positive constants and $a\thickapprox 0.618.$ The viscosity and the
heat diffusion operators $L_1,$ $L_2$ and $L_3$ are given by
\begin{align*}
L_1=&-\frac{1}{Re_1}\Delta-\frac{1}{Re_2}\frac{\partial^2}{\partial
z^2},\\
L_2=&-\frac{1}{Rt_1}\Delta-\frac{1}{Rt_2}\frac{\partial^2}{\partial
z^2},\\
L_3=&-\frac{1}{Rt_3}\Delta-\frac{1}{Rt_4}\frac{\partial^2}{\partial
z^2},
\end{align*}
where $Re_1,$ $Re_2$ are positive constants representing
the horizontal and vertical Reynolds numbers, respectively, and $
Rt_1,$ $Rt_3;$ $Rt_2,$ $Rt_4$ are positive constants which stand for the
horizontal and vertical eddy diffusivity, respectively. For the sake of simplicity, let $
\nabla =(\partial_x, \partial_y)$ be the horizontal gradient
operator and let $\Delta=\partial_x^2+\partial_y^2$ be
the horizontal Laplacian. We observe that the above system is
similar to the $3D$ Boussinesq system with the equation of
vertical motion is
approximated by the hydrostatic balance.

   Denote by $\Gamma_u$, $\Gamma_b$ and $\Gamma_l$ the upper, the bottom
   and the lateral boundaries of $\Omega$, respectively. They are
   given by
\begin{align*}
\Gamma_u=&\{(x,y,z)\in \overline{\Omega}:z=1\},\\
\Gamma_b=&\{(x,y,z)\in \overline{\Omega}:z=0\},\\
\Gamma_l=&\{(x,y,z)\in \overline{\Omega}:(x,y)\in \partial
M,0\leq z\leq 1\}.
\end{align*}
Equations \eqref{1.1}- \eqref{1.5} are subject to the following boundary conditions
\begin{align}\label{1.6}
& \frac{\partial v}{\partial
z}|_{\Gamma_u}=0,w|_{\Gamma_u}=0,(\frac{1}{Rt_2}\frac{\partial T}{\partial z}+\alpha
T)|_{\Gamma_u}=0,(\frac{1}{Rt_4}\frac{\partial q}{\partial z}+\beta
q)|_{\Gamma_u}=0,\\
\label{1.7} &\frac{\partial
v}{\partial z}|_{\Gamma_b}=0,w|_{\Gamma_b}=0,\frac{\partial T}{\partial z}|_{\Gamma_b}=0,\frac{\partial q}{\partial z}|_{\Gamma_b}=0,\\
\label{1.8}&v\cdot \vec{n}|_{\Gamma_l}=0,\frac{\partial v}{\partial
\vec{n}}\times\vec{ n}|_{\Gamma_l}=0,\frac{\partial T}{\partial \vec{n}}|_{\Gamma_l}=0,\frac{\partial q}{\partial \vec{n}}|_{\Gamma_l}=0,
\end{align}
where $ \vec{n} $ is the normal vector to $\Gamma_l,$ $\alpha,$ $\beta$ are positive constants.

 In addition, we add the initial conditions to the system \eqref{1.1}-\eqref{1.8}
\begin{align}\label{1.9}
v(x,y,z,0)=&v_0(x,y,z),\\
\label{1.10}T(x,y,z,0)=&T_0(x,y,z),\\
\label{1.11}q(x,y,z,0)=&q_0(x,y,z).
\end{align}

 In the past several decades, the primitive
equations of the atmosphere, the ocean and the coupled
atmosphere-ocean have been extensively studied from the mathematical
point of view (see \cite{ccs1, ebd, gbl2,gbl, gbl1, hc1,hc, ljl, ljl1, mtt,
pm} etc). By introducing $p$-coordinate system and using some technical treatments, Lions, Temam and Wang in \cite{ljl} obtained a new formulation for the
primitive equations of large-scale dry atmosphere which is a little similar with Navier-Stokes
equations of incompressible fluid, and they proved the existence of
weak solutions for the primitive equations of the atmosphere. In
\cite{ljl1}, Lions, Temam and Wang introduced the primitive equations of large-scale ocean and proved the existence of weak solutions and the well-posedness of local in time strong solutions for the primitive equations of large-scale ocean, and estimated the dimension of the universal attractor. Based on the works of
Lions, Temam and Wang in \cite{ljl, ljl1}, many authors continued to consider the well-posedness of solutions for the primitive equations of large-scale atmosphere (see \cite{ccs3,ccs4,ccs2, ebd, gf,gbl2, gbl3, hc, hc1, mtt, tr1,zmc1, zmc}). However, the uniqueness of weak solutions and the global existence of strong solutions for the three dimensional primitive equations of large-scale ocean and atmosphere dynamics with any initial datum remain unresolved. Until 2007, Cao and Titi \cite{ccs1} decomposed the three dimensional primitive equations of large-scale ocean and atmosphere dynamics into two systems by using the idea of the decomposition of semigroup, one is similar with the two dimensional incompressible Navier-Stokes equations, the other is the reaction-convection-diffusion equations. As we known, the solutions of each system were fairly regular. Cao and Titi performed some a priori estimates about the solutions of each system by which they obtained some a priori estimates of strong solutions for the three dimensional primitive equations of large-scale ocean and atmosphere dynamics, which implies the well-posedness of strong solutions for the three dimensional primitive equations of large-scale ocean and atmosphere dynamics, they resolved the open question posed in \cite{ljl, ljl1}. Meanwhile, the long-time behavior of solutions for the three dimensional primitive equations of large-scale ocean and atmosphere dynamics has been considered extensively (see \cite{ci, elc, gbl2,gbl, gbl1, gbl3, hng,hch, jn, jn2, jn1, lk, yb2}). In particular, in \cite{gbl}, Guo and Huang obtained a weakly compact global attractor $\mathcal{A}$ for the primitive equations of large-scale atmosphere which captures all the trajectories. The existence of a global attractor in $V$ for the primitive equations of large-scale atmosphere and ocean dynamics was proved by Ning Ju in \cite{jn} by using the Aubin-Lions compactness theorem under the assumption $Q\in L^2(\Omega).$ In \cite{jn2, jn1}, the authors have proved the finite dimensional global attractor for the 3D viscous primitive equations by using the squeezing property. As we known, the solutions of the stationary primitive equations of large-scale moist atmosphere are contained in the global attractor for the corresponding evolutionary primitive equations of large-scale moist atmosphere, it is meaningful to consider the regularity of the global attractor for the three dimensional primitive equations of large-scale moist atmosphere.

Nowadays, the study of exponential  attractors has also an interest on its own. In contrast to an exponential attractor, the global attractor has two essential drawbacks: on the one hand, the rate of attraction of the trajectories may be small and it is usually very difficult to estimate this rate in terms of the physical parameters of the problem. On the other hand, it is very sensitive to perturbations such that the global attractor can change drastically under very small perturbations of the initial dynamical system. These drawbacks obviously lead to essential difficulties in numerical simulations of global attractors and even make the global attractor unobservable in some sense. However, an exponential attractor attracts exponentially the trajectories and will thus be more stable. Furthermore, in some situations, the global attractor can be very simple and thus fails to capture interesting transient behaviors. In such situations, an exponential attractor could be a more suitable object. Therefore, it is useful to explore the existence of an exponential attractor for the three dimensional primitive equations of large-scale moist atmosphere.

The main purpose of this paper is to study the long-time behavior of solutions for the three dimensional viscous
primitive equations of large-scale moist atmosphere. In the next section, we reformulate problem \eqref{1.1}-\eqref{1.11} and give some notations used in the sequel. Section 3 is devoted to performing some a priori estimates of solutions of problem \eqref{2.4}-\eqref{2.12} to obtain the existence of absorbing sets in $V$ and $(H^2(\Omega))^4\cap V$ of the semigroup generated by problem \eqref{2.4}-\eqref{2.12}. In the last section, we prove the existence of a global attractor in $(H^2(\Omega))^4\cap V$ for problem \eqref{2.4}-\eqref{2.12} by asymptotic a priori estimate and construct an exponential attractor by using the smoothing property of the semigroup generated by problem \eqref{2.4}-\eqref{2.12}. As a byproduct, we obtain the fractal dimension of the global attractor for the semigroup generated by problem \eqref{2.4}-\eqref{2.12} is finite, which is in consistent with the results in \cite{jn1,jn2}

 Throughout this paper, let $X$ be a Banach space endowed with the norm $\|\cdot\|_X$ and let $\|u\|_p$ be the
  $L^p(\Omega)$-norm of $u$ for $1\leq p\leq\infty,$ and let $C$  be a generic positive constant.

\section{New formulation and functional setting}
\def\theequation{2.\arabic{equation}}\makeatother
\setcounter{equation}{0}
\subsection{New formulation}
    Integrating the
equation \eqref{1.3} in the $z$ direction, we obtain
\begin{align*}
w(x,y,z,t)=w(x,y,0,t)-\int_0^z\nabla\cdot v(x,y,\zeta,t)\,d\zeta.
\end{align*}

    Employing $ w(x,y,0,t)=w(x,y,1,t)=0 $ (see\eqref{1.6} and
\eqref{1.7}), we find
\begin{align}\label{2.1}
w(x,y,z,t)=-\int_0^z\nabla\cdot v(x,y,\zeta,t)\,d\zeta
\end{align}
and
\begin{align}\label{2.2}
\int_0^1\nabla\cdot
v(x,y,\zeta,t)\,d\zeta=\nabla\cdot\int_0^1
v(x,y,\zeta,t)\,d\zeta=0.
\end{align}

Integrating the equation  \eqref{1.2} with respect to $ z
$, we obtain
\begin{align}\label{2.3}
\Phi(x,y,z,t)=\Phi_s(x,y,t)-\int_0^{z}\frac{bP}{p(\zeta)}(1+aq(x,y,\zeta,t))T(x,y,\zeta,t)\,d\zeta,
\end{align}
where $ \Phi_s(x,y,t) $ is a free function to be determined.

    We infer from \eqref{2.1}, \eqref{2.3} that the following new formulation
for problem \eqref{1.1}-\eqref{1.11}
\begin{align}
\label{2.4}\nonumber&\frac{\partial v}{\partial t}+(v\cdot
\nabla)v-(\int_0^z\nabla\cdot
v(x,y,\zeta,t)\,d\zeta)\frac{\partial v}{\partial z}+\nabla
\Phi_s(x,y,t)+\frac{1}{Ro}fv^{\bot}+L_1v\\
&-\int_0^{z}\frac{bP}{p(\zeta)}\nabla [(1+aq(x,y,\zeta,t))T(x,y,\zeta,t)]\,d\zeta=0,\\
\label{2.5}&\frac{\partial T}{\partial t}+v\cdot \nabla
T-(\int_0^z\nabla\cdot v(x,y,\zeta,t)\,d\zeta)\frac{\partial T}{\partial z}+L_2T+\frac{bP}{p}(1+aq)(\int_0^z\nabla\cdot v(x,y,\zeta,t)\,d\zeta)=Q_1,\\
\label{2.6}& \frac{\partial q}{\partial t}+v\cdot\nabla
q-(\int_0^z\nabla\cdot v(x,y,\zeta,t)\,d\zeta)\frac{\partial
q}{\partial z}+L_3q=Q_2
\end{align}
with the following boundary conditions
\begin{align}
\label{2.7}&\frac{\partial v}{\partial z}|_{\Gamma_u}=0,\frac{\partial
v}{\partial z}|_{\Gamma_b}=0,v\cdot
\vec{n}|_{\Gamma_l}=0,\frac{\partial v}{\partial
\vec{n}}\times\vec{
n}|_{\Gamma_l}=0,\\
\label{2.8}&(\frac{1}{Rt_2}\frac{\partial T}{\partial z}+\alpha
T)|_{\Gamma_u}=0,\frac{\partial T}{\partial z}|_{\Gamma_b}=0,
\frac{\partial T}{\partial\vec{n}}|_{\Gamma_l}=0,\\
\label{2.9}&(\frac{1}{Rt_4}\frac{\partial q}{\partial z}+\beta
q)|_{\Gamma_u}=0,\frac{\partial q}{\partial z}|_{\Gamma_b}=0,
\frac{\partial q}{\partial\vec{n}}|_{\Gamma_l}=0
\end{align}
and the initial data
\begin{align}
\label{2.10}v(x,y,z,0)=&v_0(x,y,z),\\
\label{2.11}T(x,y,z,0)=&T_0(x,y,z),\\
\label{2.12}q(x,y,z,0)=&q_0(x,y,z).
\end{align}

 Denote
\begin{align*}
 \bar{v}(x,y)=\int_0^1v(x,y,\zeta)\,d\zeta
\end{align*}
and
\begin{align*}
  \tilde{v}=v-\bar{v}.
\end{align*}

Taking the average of \eqref{2.4} and combining Green's formula with the boundary conditions \eqref{2.7}, we obtain
\begin{align}\label{2.13}
\nonumber&\frac{\partial \bar{v}}{\partial
t}+(\bar{v}\cdot\nabla)\bar{v}+\overline{(\tilde{v}\cdot\nabla)\tilde{v}+(\nabla\cdot
\tilde{v})\tilde{v}}+\nabla
\Phi_{s}(x,y,t)-\frac{1}{Re_{1}}\Delta\bar{v}\\
&+\frac{1}{Ro}f\bar{v}^{\bot}-\overline{\int_0^{z}\frac{bP}{p(\zeta)}\nabla [(1+aq(x,y,\zeta,t))T(x,y,\zeta,t)]\,d\zeta} =0
\end{align}
which is subject to the boundary conditions
\begin{align}\label{2.14}
\nabla\cdot \bar{v}=0, \bar{v}\cdot \vec{n}|_{\Gamma_l}=0,\; \frac{\partial
\bar{v}}{\partial \vec{n}}\times \vec{n}|_{\Gamma_l}=0.
\end{align}
Substracting \eqref{2.13} from \eqref{2.4}, we have
\begin{align}\label{2.15}
\nonumber&\frac{\partial \tilde{v}}{\partial t}+(\tilde{v}\cdot
\nabla)\tilde{v}-(\int_{-h}^{z}\nabla\cdot
\tilde{v}(x,y,\zeta,t)\,d\zeta)\frac{\partial \tilde{v}}{\partial
z}+(\tilde{v}\cdot\nabla)\bar{v}+(\bar{v}\cdot\nabla)\tilde{v}-\int_0^{z}\frac{bP}{p(\zeta)}\nabla [(1+aq(x,y,\zeta,t))T(x,y,\zeta,t)]\,d\zeta\\
&+\frac{1}{Ro}f\tilde{v}^{\bot}+L_1\tilde{v}+\overline{\int_0^{z}\frac{bP}{p(\zeta)}\nabla [(1+aq(x,y,\zeta,t))T(x,y,\zeta,t)]\,d\zeta
}-\overline{(\tilde{v}\cdot\nabla)\tilde{v}+(\nabla\cdot
\tilde{v})\tilde{v}}=0,
\end{align}
which is supplemented with the boundary conditions
 \begin{align}
\label{2.16}\frac{\partial \tilde{v}}{\partial
z}|_{\Gamma_u}=0,\frac{\partial \tilde{v}}{\partial
z}|_{\Gamma_b}=0,\tilde{v}\cdot \vec{n}|_{\Gamma_l}=0,\frac{\partial
\tilde{v}}{\partial \vec{n}}\times \vec{n}|_{\Gamma_l}=0.
\end{align}
\subsection{Functional spaces and some lemmas}
To study problem \eqref{2.4}-\eqref{2.12}, we introduce some function spaces. Let
\begin{align*}
\mathcal{V}_1=&\left\{v\in (C^{\infty}(\bar{\Omega}))^2:\frac{\partial
v}{\partial z}|_{\Gamma_u}=0,\frac{\partial v}{\partial
z}|_{\Gamma_b}=0,v\cdot \vec{n}|_{\Gamma_l}=0,\frac{\partial
v}{\partial \vec{n}}\times\vec{n}|_{\Gamma_l}=0,\int_0^1\nabla\cdot v(x,y,\zeta)\,d\zeta=0\right\},\\
\mathcal{V}_2=&\left\{T\in
C^{\infty}(\bar{\Omega}):(\frac{1}{Rt_2}\frac{\partial T}{\partial
z}+\alpha T)|_{\Gamma_u}=0,\frac{\partial T}{\partial z}|_{\Gamma_b}=0,
\frac{\partial T}{\partial\vec{n}}|_{\Gamma_l}=0\right\},\\
\mathcal{V}_3=&\left\{q\in
C^{\infty}(\bar{\Omega}):(\frac{1}{Rt_4}\frac{\partial q}{\partial
z}+\beta q)|_{\Gamma_u}=0,\frac{\partial q}{\partial z}|_{\Gamma_b}=0,
\frac{\partial q}{\partial\vec{n}}|_{\Gamma_l}=0\right\}.
\end{align*}
Denote the closure of $\mathcal{V}_1,$ $\mathcal{V}_2,$ $\mathcal{V}_3$ by $V_1,$ $V_2,$ $V_3$ with respect to the
following norms, respectively, given by
\begin{align*}
\|v\|^2=&\frac{1}{Re_1}\int_{\Omega}|\nabla v|^2\,dxdydz+\frac{1}{Re_2}\int_{\Omega}|\partial_z v|^2\,dxdydz,\\
\|T\|^2=&\frac{1}{Rt_1}\int_{\Omega}|\nabla T|^2\,dxdydz+\frac{1}{Rt_2}\int_{\Omega}|\partial_z T|^2\,dxdydz+\alpha\int_M|T(z=1)|^2\,dxdy,\\
\|q\|^2=&\frac{1}{Rt_3}\int_{\Omega}|\nabla q|^2\,dxdydz+\frac{1}{Rt_4}\int_{\Omega}|\partial_z q|^2\,dxdydz+\beta\int_M|q(z=1)|^2\,dxdy,\\
\|(v,T,q)\|_V^2=&\|v\|^2+\|T\|^2+\|q\|^2,\|(v,T,q)\|_H^2=\|v\|_{L^2(\Omega)}^2+\|T\|_{L^2(\Omega)}^2+\|q\|_{L^2(\Omega)}^2
\end{align*}
for any $v\in \mathcal{V}_1,$ $T\in \mathcal{V}_2,$ $q\in \mathcal{V}_3,$ and let
 $ H_1= $ the closure of $ \mathcal{V}_1 $ with respect to the norm in $(L^2(\Omega))^2,$
$V=V_1\times V_2\times V_3,$ $H=H_1\times L^2(\Omega)\times L^2(\Omega).$

\section{Some a priori estimates of strong solutions}
\def\theequation{3.\arabic{equation}}\makeatother
\setcounter{equation}{0}
\subsection{The well-posedness of strong solutions}
 We start with the following general existence and uniqueness of
solutions for problem \eqref{2.4}-\eqref{2.12} which can be obtained by the methods used in \cite{ccs1,jn,kg,tr}. Here we only state it.
\begin{theorem}\label{3.1}
Assume that $Q_1\in L^2(\Omega)$ and $Q_2\in L^2(\Omega)$. Then for each $ (v_0, T_0, q_0)\in V,$ there exists a unique
strong solution $(v,T,q)\in C(\mathbb{R}^+;V)$ for problem \eqref{2.4}-\eqref{2.12}, which depends continuously on the initial data in $V.$
\end{theorem}

  By Theorem \ref{3.1}, we can define the operator semigroup $\{S(t)\}_{t\geq 0}$ in $V$ as
\begin{align*}
S(.): \mathbb{R}^+\times V\rightarrow V,
\end{align*}
which is $(V,V)$-continuous.
\subsection{Some a priori estimates of strong solutions}
In this subsection, we give some a priori estimates of strong solutions for problem \eqref{2.4}-\eqref{2.12}, which imply the existence of absorbing sets for the semigroup $\{S(t)\}_{t\geq 0}$ associated with problem \eqref{2.4}-\eqref{2.12}.
\subsubsection{$L^2(\Omega)$ estimates of $q$}
 Taking the inner product of \eqref{2.6} with $q$ in $L^2(\Omega),$ we obtain
\begin{align}\label{3.1.1}
\frac{1}{2}\frac{d}{dt}\|q\|^2_2+\|q\|^2=\int_{\Omega}Q_2q\,dxdydz.
\end{align}
Thanks to
\begin{align*}
\|q\|^2_2\leq 2\|q(z=1)\|^2_{L^2(M)}+2\|\partial_zq\|^2_2,
\end{align*}
we find
\begin{align}\label{3.1.2}
\frac{\|q\|^2_2}{2Rt_4+\frac{2}{\beta}}
\leq\frac{1}{Rt_4}\int_{\Omega}|\partial_z q|^2\,dxdydz+\beta\int_{M}|q(z=1)|^2\,dxdy.
\end{align}
It follows from \eqref{3.1.1}-\eqref{3.1.2} that
\begin{align*}
\frac{d}{dt}\|q\|^2_2+\|q\|^2\leq (2Rt_4+\frac{2}{\beta})\|Q_2\|^2_2.
\end{align*}
Using \eqref{3.1.2} again, we obtain
\begin{align*}
\frac{d}{dt}\|q\|^2_2+\frac{\|q\|^2_2}{2Rt_4+\frac{2}{\beta}} \leq
(2Rt_4+\frac{2}{\beta})\|Q_2\|^2_2.
\end{align*}
We infer from the classical Gronwall inequality that
\begin{align*}
\|q\|_2^2\leq \|q_0\|_2^2\exp(\frac{-t}{2Rt_4+\frac{2}{\beta}})+(2Rt_4+\frac{2}{\beta})^2\|Q_2\|_2^2,
\end{align*}
which implies that
\begin{align}\label{3.1.3}
\|q\|_2^2+\int_t^{t+1}\|q(\tau)\|^2\,d\tau\leq&\rho_1
\end{align}
for any $t\geq T_1$. For brevity, we omit writing out these bounds explicitly here and we also omit writing out other
similar bounds in our future discussion for all other uniform a priori estimates.
\subsubsection{$(L^2(\Omega))^3$ estimates of $(v,T)$}
 Multiplying \eqref{2.4} by $v$ and integrating over $\Omega,$ we obtain
\begin{align}\label{3.2.4}
\frac{1}{2}\frac{d}{dt}\|v\|^2_2+\|v\|^2=\int_{\Omega}\left(\int_0^{z}\frac{bP}{p(\zeta)}\nabla [(1+aq(x,y,\zeta,t))T(x,y,\zeta,t)]\,d\zeta\right)\cdot v\,dxdydz.
\end{align}
Taking the inner product of \eqref{2.5} with $T$ in $L^2(\Omega),$ we find
\begin{align}\label{3.2.5}
\frac{1}{2}\frac{d}{dt}\|T\|^2_2+\|T\|^2=\int_{\Omega}Q_1T\,dxdydz-\int_{\Omega}\left(\frac{bP}{p}(1+aq)(\int_0^z\nabla\cdot v(x,y,\zeta,t)\,d\zeta)\right)T\,dxdydz.
\end{align}
Integrating by parts and combining \eqref{2.2} with \eqref{2.7}, we obtain
\begin{align}\label{3.2.6}
\nonumber&\int_{\Omega}\left(\int_0^{z}\frac{bP}{p(\zeta)}\nabla [(1+aq(x,y,\zeta,t))T(x,y,\zeta,t)]\,d\zeta\right)\cdot v\,dxdydz\\
\nonumber=&-\int_{\Omega}\left(\int_0^{z}\frac{bP}{p(\zeta)}(1+aq(x,y,\zeta,t))T(x,y,\zeta,t)\,d\zeta\right)(\nabla\cdot v)\,dxdydz\\
=&\int_{\Omega}\left(\frac{bP}{p}(1+aq)(\int_0^z\nabla\cdot v(x,y,\zeta,t)\,d\zeta)\right)T\,dxdydz.
\end{align}
It follows from \eqref{3.2.4}-\eqref{3.2.6} and H\"{o}lder inequality  that
\begin{align}\label{3.2.7}
\frac{1}{2}\frac{d}{dt}(\|v\|^2_2+\|T\|_2^2)+\|v\|^2+\|T\|^2\leq \|Q_1\|_2\|T\|_2.
\end{align}
Notice that
\begin{align}\label{3.2.8}
\frac{\|v\|_2^2}{C_M}+\frac{\|T\|^2_2}{2Rt_2+\frac{2}{\alpha}}\leq&\|T\|^2+\|v\|^2.
\end{align}
Therefore, we deduce from \eqref{3.2.7}-\eqref{3.2.8}, H\"{o}lder inequality and Young inequality that
\begin{align*}
\frac{d}{dt}(\|v\|^2_2+\|T\|_2^2)+\frac{\|v\|^2_2}{C_M}+\frac{\|T\|^2_2}{2Rt_2+\frac{2}{\alpha}}\leq(2Rt_2+\frac{2}{\alpha})\|Q_1\|^2_2,
\end{align*}
which implies that
\begin{align}\label{3.2.9}
\|v\|^2_2+\|T\|_2^2+\int_{t}^{t+1}\|v(\tau)\|^2+\|T(\tau)\|^2\,d\tau\leq\rho_2
\end{align}
for any $t\geq T_2\geq T_1$.
\subsubsection{$L^6(\Omega)$ estimates of $q$}
Multiplying \eqref{2.6} by $|q|^4q$ and integrating over $\Omega,$ we have
\begin{align*}
\frac{1}{6}\frac{d}{dt}\|q\|^6_6+\frac{5}{9}\||q|^3\|^2
\leq& \||q|^3\|^{\frac{5}{3}}_{\frac{10}{3}}\|Q_2\|_2\\
\leq& C\|Q_2\|_2\||q|^3\|^{\frac{2}{3}}_2\||q|^3\|\\
=& C\|Q_2\|_2\|q\|^2_6\||q|^3\|.
\end{align*}
Using Young inequality, we obtain
\begin{align*}
\frac{d}{dt}\|q\|_6^2\leq C\|Q_2\|_2^2.
\end{align*}
Therefore,  we infer from the uniform Gronwall inequality and \eqref{3.2.4} that
\begin{align}\label{3.3.1}
\|q\|_6^2+\int_t^{t+1}\||q(\tau)|^3\|^2\,d\tau\leq\rho_3
\end{align}
for any $t\geq T_2+1.$
\subsubsection{$L^6(\Omega)$ estimates of $T$}
Taking the inner product of \eqref{2.5} with $|T|^4T$ in $L^2(\Omega),$ we deduce
\begin{align}\label{3.4.1}
\nonumber\frac{1}{6}\frac{d}{dt}\|T\|^6_6+\frac{5}{9}\||T|^3\|^2\leq& \||T|^3\|^{\frac{5}{3}}_{\frac{10}{3}}\|Q_1\|_2+\int_{\Omega}\frac{bP}{p}(1+aq)(\int_0^z\nabla\cdot v(x,y,\zeta,t)\,d\zeta)|T|^4T\,dxdydz\\
\leq& C\|Q_1\|_2\||T|^3\|^{\frac{2}{3}}_2\||T|^3\|+C\|\nabla v\|_2\||T|^3\|^{\frac{2}{3}}_2\||T|^3\|+I_1,
\end{align}
where
\begin{align*}
I_1=\int_{\Omega}\frac{abP}{p}q(\int_0^z\nabla\cdot v(x,y,\zeta,t)\,d\zeta)|T|^4T\,dxdydz.
\end{align*}
Now, we estimate $I_1$ as follows.
\begin{align}\label{3.4.2}
\nonumber I_1\leq&C\int_0^1\|q\|_{L^6(M)}\left\|\int_0^1|\nabla v|(x,y,\zeta,t)\,d\zeta\right\|_{L^2(M)}\||T|^5\|_{L^3(M)}\,dz\\
\nonumber\leq&C\left\|\int_0^1|\nabla v|(x,y,\zeta,t)\,d\zeta\right\|_{L^2(M)}\int_0^1\|q\|_{L^6(M)}\||T|^3\|_{L^2(M)}^{\frac{2}{3}}\||T|^3\|_{H^1(M)}\,dz\\
\nonumber\leq&C\left(\int_0^1\|\nabla v\|_{L^2(M)}\,d\zeta\right)\|q\|_6\||T|^3\|_2^{\frac{2}{3}}\||T|^3\|\\
\leq&C\|\nabla v\|_2\|q\|_6\||T|^3\|_2^{\frac{2}{3}}\||T|^3\|.
\end{align}
We deduce from \eqref{3.4.1}-\eqref{3.4.2} that
\begin{align*}
\frac{d}{dt}\|T\|^2_6
\leq C\|Q_1\|_2^2+C\|\nabla v\|_2^2+C\|q\|_6^2\|\nabla v\|_2^2.
\end{align*}
Combining the uniform Gronwall inequality with \eqref{3.2.9}, \eqref{3.3.1}, we obtain
\begin{align}\label{3.4.3}
\|T\|_6^2+\int_t^{t+1}\||T(\tau)|^3\|^2\,d\tau\leq\rho_4
\end{align}
for any $t\geq T_2+2.$
\subsubsection{$(L^6(\Omega))^{2}$ estimates of $\tilde{v}$}
Multiplying \eqref{2.15} by $|\tilde{v}|^4\tilde{v}$ and integrating over $\Omega,$ we deduce
\begin{align}\label{3.5.1}
\nonumber&\frac{1}{6}\frac{d}{dt}\|
\tilde{v}\|^6_6+\frac{1}{Re_1}\int_{\Omega}|\nabla
\tilde{v}|^2|\tilde{v}|^4\,dxdydz+\frac{1}{Re_2}\int_{\Omega}|\partial_z
\tilde{v}|^2|\tilde{v}|^4\,dxdydz+\frac{4}{9}\||\tilde{v}|^3\|^2\\
\leq& C\int_{\Omega}|\bar{v}||\nabla
\tilde{v}||\tilde{v}|^5\,dxdydz+C\int_M(\int_0^1|\tilde{v}|^2\,dz)(\int_0^1|\nabla
\tilde{v}||\tilde{v}|^4\,dz)\,dxdy+I_2,
\end{align}
where
\begin{align*}
I_2=\int_{\Omega}\left(\int_0^z\frac{bP}{p(\zeta)}[(1+aq)T]\,d\zeta-\int_0^1\int_0^{\eta}\frac{bP}{p(\zeta)} [(1+aq)T]\,d\zeta\,d\eta
\right)(\nabla\cdot |\tilde{v}|^4\tilde{v})\,dxdydz.
\end{align*}
In the following, we estimate $I_2$ by using H\"{o}lder inequality.
\begin{align*}
 I_2\leq&C\left\|\int_0^z\frac{bP}{p(\zeta)}[(1+aq)T]\,d\zeta\right\|_6\||\nabla \tilde{v}||\tilde{v}|^2\|_2\|\tilde{v}\|_6^2\\
\leq&C\|T\|_6\||\nabla \tilde{v}||\tilde{v}|^2\|_2\|\tilde{v}\|_6^2+C\left\|\int_0^1|qT|\,d\zeta\right\|_{L^6(M)}\||\nabla \tilde{v}||\tilde{v}|^2\|_2\|\tilde{v}\|_6^2.
\end{align*}
Due to
\begin{align*}
\left\|\int_0^1|qT|\,d\zeta\right\|_{L^6(M)}^6=&\int_M\left|\int_0^1|qT|\,dz\right|^6\,dxdy\\
\leq&\int_M(\int_0^1|q|^2\,dz)^3(\int_0^1|T|^2\,dz)^3\,dxdy\\
\leq&\left(\int_M(\int_0^1|q|^2\,dz)^6\,dxdy\right)^{\frac{1}{2}}\left(\int_M(\int_0^1|T|^2\,dz)^6\,dxdy\right)^{\frac{1}{2}}\\
\leq&\left(\int_0^1(\int_M|q|^{12}\,dxdy)^{\frac{1}{6}}\,dz\right)^3\left(\int_0^1(\int_M|T|^{12}\,dxdy)^{\frac{1}{6}}\,dz\right)^3\\
\leq&C\left(\int_0^1\|q\|_{L^6(M)}\|q\|_{H^1(M)}\,dz\right)^3\left(\int_0^1\|T\|_{L^6(M)}\|T\|_{H^1(M)}\,dz\right)^3\\
\leq&C\|q\|_6^3\|q\|^3\|T\|_6^3\|T\|^3,
\end{align*}
which implies that
\begin{align}\label{3.5.2}
I_2\leq C\|T\|_6\|\nabla \tilde{v}||\tilde{v}|^2\|_2\|\tilde{v}\|_6^2+C\|q\|_6^{\frac{1}{2}}\|q\|^{\frac{1}{2}}\|T\|_6^{\frac{1}{2}}\|T\|^{\frac{1}{2}}\||\nabla \tilde{v}||\tilde{v}|^2\|_2\|\tilde{v}\|_6^2.
\end{align}
It follows from H\"{o}lder inequality that
\begin{align}\label{3.5.3}
\nonumber\int_{\Omega}|\bar{v}||\nabla
\tilde{v}||\tilde{v}|^5\,dxdydz\leq&\int_M|\bar{v}|(\int_0^1|\nabla
\tilde{v}|^2|\tilde{v}|^4\,dz)^{\frac{1}{2}}(\int_0^1|
\tilde{v}|^6\,dz)^{\frac{1}{2}}\,dxdy\\
\leq &\|\bar{v}\|_{L^4(M)}\||\nabla
\tilde{v}||\tilde{v}|^2\|_2(\int_0^1(\int_{M}|\tilde{v}|^{12}\,dxdy)^{\frac{1}{2}}\,dz)^{\frac{1}{2}}.
\end{align}
Thanks to
\begin{align*}
\int_M|\tilde{v}|^{12}\,dxdy=&\int_{M}||\tilde{v}|^3|^4\,dxdy\\
\leq&
C\int_M|\tilde{v}|^6\,dxdy\int_M|\nabla|\tilde{v}|^3|^2\,dxdy,
\end{align*}
we obtain
\begin{align}\label{3.5.4}
(\int_0^1(\int_M|\tilde{v}|^{12}\,dxdy)^{\frac{1}{2}}\,dz)^{\frac{1}{2}}
\leq
C(\int_{\Omega}|\tilde{v}|^6\,dxdydz)^{\frac{1}{4}}(\int_{\Omega}|\nabla|\tilde{v}|^3|^2\,dxdydz)^{\frac{1}{4}}.
\end{align}
Therefore,we deduce from \eqref{3.5.3}-\eqref{3.5.4} that
\begin{align}\label{3.5.5}
\int_{\Omega}|\bar{v}||\nabla \tilde{v}||\tilde{v}|^5\,dxdydz\leq
C\|\tilde{v}\|_6^{\frac{3}{2}}\|v\|_2^{\frac{1}{2}}\|\nabla v\|_2^{\frac{1}{2}}(\int_{\Omega}|\nabla|\tilde{v}|^3|^2\,dxdydz)^{\frac{1}{4}}
(\int_{\Omega}|\nabla
\tilde{v}|^{2}|\tilde{v}|^4\,dxdydz)^{\frac{1}{2}}.
\end{align}
Repeating the similar process with the above, we deduce
\begin{align}\label{3.5.6}
\int_{M}(\int_0^1|\tilde{v}|^{2}\,dz)(\int_0^1|\nabla
\tilde{v}||\tilde{v}|^4\,dz)\,dxdy\leq C\||\nabla
\tilde{v}||\tilde{v}|^2\|_2\|
\tilde{v}\|^3_6\|\tilde{v}\|_{H^{1}(\Omega)}.
\end{align}
 We infer from \eqref{3.5.1}-\eqref{3.5.2}, \eqref{3.5.5}-\eqref{3.5.6} that
\begin{align*}
&\frac{d}{dt}\|\tilde{v}\|^6_6+\frac{2}{Re_1}\int_{\Omega}|\nabla\tilde{v}|^2|\tilde{v}|^4\,dxdydz+\frac{2}{Re_2}\int_{\Omega}|\partial_{z}
\tilde{v}|^2|\tilde{v}|^4\,dxdydz+2\||\tilde{v}|^3\|^2\\
\leq& C(\|v\|^2_2\|\nabla v\|^2_2+\|\tilde{v}\|^2_{H^1(\Omega)})\|\tilde{v}\|^6_6+C\|T\|_6^2\|\tilde{v}\|_6^4+C\|q\|_6\|q\|\|T\|_6\|T\|\|\tilde{v}\|_6^4.
\end{align*}
Therefore, it follows from \eqref{3.1.3}, \eqref{3.2.9}, \eqref{3.3.1} and \eqref{3.4.3} that
\begin{align}\label{3.5.7}
\|\tilde{v}\|^2_6+\int_t^{t+1}\int_{\Omega}|\nabla
\tilde{v}|^2|\tilde{v}|^4\,dxdydz\,d\tau\leq\rho_5
\end{align}
for any $t\geq T_2+3.$
\subsubsection{$(H^{1}(M))^{2}$ estimates of $\bar{v}$}
 Taking the inner product of equation \eqref{2.13} with
$-\Delta \bar{v}$ in $L^{2}(\Omega)$ and combining the boundary conditions
\eqref{2.14}, we obtain
\begin{align}\label{3.6.1}
\frac{1}{2}\frac{d}{dt}\|
\nabla\bar{v}\|^{2}_{L^{2}(M)}+\frac{1}{Re_{1}}\int_{M}|\Delta
\bar{v}|^{2}\,dxdy\leq C\int_{M}|\bar{v}||\nabla \bar{v}||\Delta
\bar{v}|\,dxdy +C\int_{M}(\int_0^1|\nabla
\tilde{v}||\tilde{v}|\,dz)|\Delta \bar{v}|\,dxdy,
\end{align}
where we have used the following equalities
\begin{align*}
&\int_M\nabla\Phi_{s}(x,y,t)\cdot\Delta\bar{v}\,dxdy=0,\\
&\frac{1}{Ro}\int_Mf\bar{v}^{\bot}\cdot\Delta\bar{v}\,dxdy=0,\\
&\int_M\overline{\int_0^{z}\frac{bP}{p(\zeta)}\nabla [(1+aq(x,y,\zeta,t))T(x,y,\zeta,t)]\,d\zeta}\cdot\Delta\bar{v}\,dxdy=0.
\end{align*}
In the following, we give the estimates of each term of the right hand side of \eqref{3.6.1}.
\begin{align}
\label{3.6.2}\nonumber\int_{M}|\bar{v}||\nabla \bar{v}||\Delta\bar{v}|\,dxdy
\leq&C\|\bar{v}\|_{L^4(M)}\|\nabla\bar{v}\|_{L^4(M)}\|\Delta\bar{v}\|_{L^2(M)}\\
\leq& C\|\bar{v}\|_{L^2(M)}^{\frac{1}{2}}
\|\nabla\bar{v}\|_{L^2(M)}\|\Delta\bar{v}\|_{L^2(M)}^{\frac{3}{2}},\\
\label{3.6.3} \int_{M}(\int_0^1|\tilde{v}||\nabla
\tilde{v}|\,dz)|\Delta \bar{v}|\,dxdy
\leq &C\||\nabla
\tilde{v}||\tilde{v}|^2\|_2^{\frac{1}{2}}\|\nabla\tilde{v}\|_2^{\frac{1}{2}}\|\Delta\bar{v}\|.
\end{align}
It follows from \eqref{3.6.1}-\eqref{3.6.3} that
\begin{align*}
\frac{d}{dt}\|
\nabla\bar{v}\|^2_{L^2(M)}+\frac{1}{Re_1}\int_{M}|\Delta
\bar{v}|^2\,dxdy
\leq C\| \bar{v}\|^2_{L^2(M)}\|\nabla\bar{v}\|_{L^2(M)}^4+C\||\nabla
\tilde{v}||\tilde{v}|^2\|_2^2+C\|\nabla\tilde{v}\|_2^2.
\end{align*}
In view of \eqref{3.2.9}, \eqref{3.5.7} and the uniform Gronwall inequality, we obtain
\begin{align}\label{3.6.4}
\|\nabla\bar{v}\|^{2}_{L^2(M)}\leq\rho_6
\end{align}
for any $t\geq T_2+4.$
\subsubsection{$(L^2(\Omega))^2$ estimates of $v_z$}
 Denoted by $u=v_z$. It is clear that $u$ satisfies the following
 equation obtained by differentiating the equation \eqref{2.4} with respect to
 $z$:
 \begin{align}\label{3.7.1}
\frac{\partial u}{\partial t}+L_1u+(v\cdot\nabla)u-(\int_0^z\nabla\cdot v(x,y,\zeta,t)d\zeta)\frac{\partial u}{\partial z}+(u\cdot\nabla)v-(\nabla\cdot v)u+\frac{1}{Ro}fu^{\bot}-\frac{bP}{p}\nabla [(1+aq)T]=0
 \end{align}
subject to the boundary conditions
 \begin{align}
\label{3.7.2}u|_{\Gamma_u}=0,u|_{\Gamma_b}=0,u\cdot \vec{n}|_{\Gamma_l}=0,\frac{\partial
u}{\partial \vec{n}}\times \vec{n}|_{\Gamma_l}=0.
\end{align}
Multiplying \eqref{3.7.1} by $u$ and integrating over $\Omega,$ we find
\begin{align}\label{3.7.3}
\nonumber\frac{1}{2}\frac{d}{dt}\|u\|^2_2+\|u\|^2=&-\int_{\Omega}[(u\cdot\nabla)v-(\nabla\cdot v)u -\frac{bP}{p}\nabla \left((1+aq)T\right)]\cdot u\,dxdydz\\
\leq& C\int_{\Omega}|v||u||\nabla u|\,dxdydz +C\int_{\Omega}|T||\nabla u|\,dxdydz+C\int_{\Omega}|T||q||\nabla u|\,dxdydz.
\end{align}
Next, we estimate the right hand side of \eqref{3.7.3} term by term.
\begin{align}
\label{3.7.4}\int_{\Omega}|T||\nabla u|\,dxdydz\leq& \|T\|_{2}\|\nabla u\|_{2},\\
\label{3.7.5}\nonumber \int_{\Omega}|v||u||\nabla u|\,dxdydz\leq&\|v\|_6\| u\|_3\|\nabla u\|_2\\
\leq& C\|v\|_6\|u\|^{\frac{1}{2}}_2\|u\|^{\frac{3}{2}},\\
\label{3.7.6}\nonumber\int_{\Omega}|T||q||\nabla u|\,dxdydz\leq &\|T\|_3\|q\|_6\|\nabla u\|_2\\
\leq &C\|T\|_2^{\frac{1}{2}}\|T\|^{\frac{1}{2}}\|q\|_6\|\nabla u\|_2.
\end{align}
\indent
It follows from \eqref{3.7.3}-\eqref{3.7.6} that
\begin{align*}
\frac{d}{dt}\|u\|^2_2+\|u\|^2\leq C\|v\|_6^4\|u\|^2_2+C\|T\|^2_2+C\|q\|_6^4\|T\|^2.
\end{align*}

It is shown in \cite{jn} that
\begin{align*}
\|v\|_6\leq C\|v\|_2+C\|\nabla \bar{v}\|_2+\|\tilde{v}\|_6,
\end{align*}
which implies that
\begin{align}\label{3.7.7}
\|v\|_6^2\leq \rho_7
\end{align}
for any $t\geq T_2+4.$

 Thanks to the uniform Gronwall inequality, \eqref{3.2.9} and
\eqref{3.7.7}, we obtain
\begin{align}\label{3.7.8}
\|\partial_{z}v\|^2_2+\int_{t}^{t+1}\|\partial_{z}v(\tau)\|^2\,d\tau\leq\rho_8
\end{align}
for any $t\geq T_2+5$.
\subsubsection{$(L^2(\Omega))^2$ estimates of $(T_z,q_z)$}
Taking the inner product of equation
\eqref{2.6} with $-\frac{\partial^2 q}{\partial z^2}$ in $L^2(\Omega)$ and combining the boundary
conditions \eqref{2.2}, \eqref{2.9}, we find
\begin{align}\label{3.8.1}
\nonumber&\frac{1}{2}\frac{d}{dt}(\|q_z\|^2+Rt_4\beta\|q\|_{L^2(\Gamma_u)}^2)+\frac{1}{Rt_3}\|\nabla q_z\|_2^2+\frac{1}{Rt_4}\|\partial_z q_z\|_2^2+\frac{\beta Rt_4}{Rt_3}\|\nabla q\|_{L^2(\Gamma_u)}^2\\
\nonumber=&-\int_{\Omega}Q_2\partial_z q_z+\int_{\Omega}\left[v\cdot\nabla q-\left(\int_0^z\nabla\cdot v(x,y,\zeta,t)\,d\zeta\right)q_z\right]\frac{\partial^2 q}{\partial z^2}\,dxdydz\\
\nonumber\leq&\|Q_2\|_2\|\partial_z q_z\|_2-Rt_4\beta\int_{\Gamma_u}(v\cdot\nabla q)q-\int_{\Omega}\left[v_z\cdot\nabla q-(\nabla\cdot v)q_z\right]q_z\,dxdydz\\
\leq&\|Q_2\|_2\|\partial_z q_z\|_2+\frac{Rt_4\beta}{2}\int_{\Gamma_u}(\nabla\cdot v)|q|^2+C\|\nabla v_z\|_2\|q\|_6\|q_z\|_3+C\|v\|_6\|q_z\|_3\|\nabla q_z\|_2+C\|v_z\|_3\|q\|_6\|\nabla q_z\|_2.
\end{align}
Using H\"{o}lder inequality, we have
\begin{align}\label{3.8.2}
\nonumber\int_{\Gamma_u}(\nabla\cdot v)|q|^2\,dxdy=&\int_M(\int_{\eta}^1\nabla\cdot v_{\zeta}(x,y,\zeta,t)\,d\zeta+\int_0^1\nabla\cdot v(x,y,\zeta,t)\,d\zeta)|q(z=1)|^2\,dxdy\\
\nonumber\leq&C(\|\nabla v_z\|_2+\|\nabla v\|_2)\|q\|_{L^4(\Gamma_u)}^2\\
\leq&C(\|\nabla v_z\|_2+\|\nabla v\|_2)\|q\|_{L^2(\Gamma_u)}\|q\|_{L^6(\Gamma_u)}^3.
\end{align}
Multiplying \eqref{2.5} by $-\frac{\partial^2 T}{\partial z^2}$ and integrating over $\Omega,$ and using the boundary conditions \eqref{2.2}, \eqref{2.8}, we find
\begin{align}\label{3.8.3}
\nonumber&\frac{1}{2}\frac{d}{dt}(\|T_z\|^2+Rt_2\alpha\|T\|_{L^2(\Gamma_u)}^2)+\frac{1}{Rt_1}\|\nabla T_z\|_2^2+\frac{1}{Rt_2}\|\partial_z T_z\|_2^2+\frac{\alpha Rt_2}{Rt_1}\|\nabla T\|_{L^2(\Gamma_u)}^2\\
\nonumber=&-\int_{\Omega}Q_1\partial_z T_z+\int_{\Omega}\left[v\cdot\nabla T-\left(\int_0^z\nabla\cdot v(x,y,\zeta,t)\,d\zeta\right)T_z\right]\frac{\partial^2 T}{\partial z^2}\,dxdydz+\int_{\Omega}\frac{bP}{p}(1+aq)(\int_0^z\nabla\cdot v(x,y,\zeta,t)\,d\zeta)\frac{\partial^2 T}{\partial z^2}\,dxdydz\\
\nonumber\leq&\|Q_1\|_2\|\partial_z T_z\|_2-Rt_2\alpha\int_{\Gamma_u}(v\cdot\nabla T)T-\int_{\Omega}\left[v_z\cdot\nabla T-(\nabla\cdot v)T_z\right]T_z\,dxdydz-\int_{\Omega}\frac{bP}{p}(1+aq)(\nabla\cdot v)T_z\,dxdydz\\
\nonumber&-\int_{\Omega}\frac{abP}{p}q_z(\int_0^z\nabla\cdot v(x,y,\zeta,t)\,d\zeta)T_z\,dxdydz+\int_{\Omega}\frac{bP(P-p_0)}{p^2}(1+aq)(\int_0^z\nabla\cdot v(x,y,\zeta,t)\,d\zeta)T_z\,dxdydz\\
\nonumber\leq&\|Q_1\|_2\|\partial_z T_z\|_2+\frac{Rt_2\alpha}{2}\int_{\Gamma_u}(\nabla\cdot v)|T|^2+C\|\nabla v_z\|_2\|T\|_6\|T_z\|_3+C\|v\|_6\|T_z\|_3\|\nabla T_z\|_2+C\|v_z\|_3\|T\|_6\|\nabla T_z\|_2\\
&+C\|q\|_6\|\nabla v\|_2\|T_z\|_3+C\|\nabla v\|_2\|T_z\|_2+C\|\nabla q_z\|_2\|v\|_6\|T_z\|_3+C\|\nabla T_z\|_2\|v\|_6\|q_z\|_3.
\end{align}
Similarly, we have
\begin{align}\label{3.8.4}
\int_{\Gamma_u}(\nabla\cdot v)|T|^2\,dxdy
\leq C(\|\nabla v_z\|_2+\|\nabla v\|_2)\|T\|_{L^2(\Gamma_u)}\|T\|_{L^6(\Gamma_u)}^3.
\end{align}
Therefore, by virtue of Young inequality, the uniform Gronwall inequality and \eqref{3.8.1}-\eqref{3.8.4}, we obtain
\begin{align}
\nonumber\label{3.8.5}&\|q_z\|^2+Rt_4\beta\|q\|_{L^2(\Gamma_u)}^2+\|T_z\|^2+Rt_2\alpha\|T\|_{L^2(\Gamma_u)}^2+\frac{1}{Rt_3}\int_t^{t+1}\|\nabla q_z\|_2^2+\frac{1}{Rt_4}\int_t^{t+1}\|\partial_z q_z\|_2^2\\
&+\frac{1}{Rt_1}\int_t^{t+1}\|\nabla T_z\|_2^2+\frac{1}{Rt_2}\int_t^{t+1}\|\partial_z T_z\|_2^2+\frac{\alpha Rt_2}{Rt_1}\int_t^{t+1}\|\nabla T\|_{L^2(\Gamma_u)}^2+\frac{\beta Rt_4}{Rt_3}\int_t^{t+1}\|\nabla q\|_{L^2(\Gamma_u)}^2\leq\rho_9
\end{align}
for any $t\geq T_2+6$.
\subsubsection{$L^2(\Omega)$ estimates of $(\nabla v, \nabla T, \nabla q)$}
Taking the inner product of equation
\eqref{2.4} with $-\Delta v$ in $L^2(\Omega)$ and combining the boundary
condition \eqref{2.7}, we have
\begin{align}\label{3.9.1}
\nonumber&\frac{1}{2}\frac{d}{dt}\|\nabla v\|^{2}_{2}+\frac{1}{Re_1}\int_{\Omega}|\Delta
v|^2\,dxdydz+\frac{1}{Re_2}\int_{\Omega}|\nabla\partial_{z} v|^2\,dxdydz\\
\nonumber\leq&C\int_M(\int_0^1|\nabla v|\,dz)(\int_0^1|\partial_z v||\Delta v|\,dz)\,dxdy+C\int_{\Omega}|v||\nabla v||\Delta v|\,dxdydz\\
&-\int_{\Omega}(\int_0^{z}\frac{bP}{p(\zeta)}\nabla [(1+aq(x,y,\zeta,t))T(x,y,\zeta,t)]\,d\zeta)\cdot\Delta v\,dxdydz.
\end{align}
In the following, we estimate each term of the right hand side of \eqref{3.9.1}.
\begin{align}\label{3.9.2}
\int_{\Omega}|v||\nabla v||\Delta v|\,dxdydz
\nonumber\leq &C\|v\|_6\| \nabla v\|_3\| \Delta v\|_2\\
\leq &C\|v\|_6\|\nabla
v\|^{\frac{1}{2}}_2(\|\nabla
\partial_{z}v\|_2+\|\Delta
v\|_2)^{\frac{3}{2}},
\end{align}
\begin{align}
\label{3.9.3}\int_{M}(\int_0^1|\nabla v|\,dz)(\int_0^1|\partial_z
v||\Delta v|\,dz)\,dxdy
\leq &C\|v_z\|_2^{\frac{1}{2}}\|\nabla v_z\|_2^{\frac{1}{2}}\|\nabla v\|_2^{\frac{1}{2}}\|\Delta v\|_2^{\frac{3}{2}}
\end{align}
and
\begin{align}\label{3.9.4}
\nonumber&\int_{\Omega}(\int_0^{z}\frac{bP}{p(\zeta)}\nabla [(1+aq(x,y,\zeta,t))T(x,y,\zeta,t)]\,d\zeta)\cdot\Delta v\,dxdydz\\
\leq&C\|\nabla T\|_2\|\Delta v\|_2+C\|q\|_6\|\nabla T\|_3\|\Delta v\|_2+C\|T\|_6\|\nabla q\|_3\|\Delta v\|_2.
\end{align}
Multiplying \eqref{2.6} by $-\Delta q$ and integrating over $\Omega,$ we have
\begin{align}\label{3.9.5}
\nonumber&\frac{1}{2}\frac{d}{dt}\|\nabla q\|^{2}_{2}+\frac{1}{Rt_3}\int_{\Omega}|\Delta
q|^2\,dxdydz+\frac{1}{Rt_4}\int_{\Omega}|\nabla\partial_z q|^2\,dxdydz+\beta\int_{\Gamma_u}|\nabla q|^2\,dxdy\\
\nonumber\leq&C\int_M(\int_0^1|\nabla v|\,dz)(\int_0^1|\partial_z q||\Delta q|\,dz)\,dxdy+C\int_{\Omega}|v||\nabla q||\Delta q|\,dxdydz+\|Q_2\|_2\|\Delta q\|_2\\
\leq&C\|\nabla v\|_2^{\frac{1}{2}}\|\Delta v\|_2^{\frac{1}{2}}\|q_z\|_2^{\frac{1}{2}}\|\nabla q_z\|_2^{\frac{1}{2}}\|\Delta q\|_2+C\|v\|_6\|\nabla q\|_3\|\Delta q\|_2+\|Q_2\|_2\|\Delta q\|_2.
\end{align}
Taking the inner product of equation
\eqref{2.5} with $-\Delta T$ in $L^2(\Omega)$ and combining the boundary
condition \eqref{2.8}, we find
\begin{align}\label{3.9.6}
\nonumber&\frac{1}{2}\frac{d}{dt}\|\nabla T\|^{2}_{2}+\frac{1}{Rt_1}\int_{\Omega}|\Delta
T|^2\,dxdydz+\frac{1}{Rt_2}\int_{\Omega}|\nabla\partial_z T|^2\,dxdydz+\alpha\int_{\Gamma_u}|\nabla T|^2\,dxdy\\
\nonumber\leq&C\int_M(\int_0^1|\nabla v|\,dz)(\int_0^1|\partial_z T||\Delta T|\,dz)\,dxdy+C\int_{\Omega}|v||\nabla T||\Delta T|\,dxdydz+\|Q_1\|_2\|\Delta T\|_2\\
\nonumber&-\int_{\Omega}\frac{bP}{p}(1+aq)(\int_0^z\nabla\cdot v(x,y,\zeta,t)\,d\zeta)\Delta T\,dxdydz\\
\nonumber\leq&C\|\nabla v\|_2^{\frac{1}{2}}\|\Delta v\|_2^{\frac{1}{2}}\|T_z\|_2^{\frac{1}{2}}\|\nabla T_z\|_2^{\frac{1}{2}}\|\Delta T\|_2+C\|v\|_6\|\nabla T\|_3\|\Delta T\|_2+\|Q_1\|_2\|\Delta T\|_2\\
&+C\|\nabla v\|_2\|\Delta T\|_2+C\|q\|_6\|\nabla v\|_3\|\Delta T\|_2.
\end{align}
From the uniform Gronwall inequality, Young inequality and \eqref{3.8.5}-\eqref{3.9.6}, we obtain
\begin{align}
\nonumber\label{3.9.7}&\|\nabla v\|_2^2+\|\nabla q\|_2^2+\|\nabla T\|_2^2+\frac{1}{Rt_3}\int_t^{t+1}\|\Delta
q\|_2^2+\frac{1}{Rt_4}\int_t^{t+1}\|\nabla\partial_z q\|_2^2+\beta\int_t^{t+1}\|\nabla q\|_{L^2(\Gamma_u)}^2+\frac{1}{Rt_1}\int_t^{t+1}\|\Delta T\|_2^2\\
&+\frac{1}{Rt_2}\int_t^{t+1}\|\nabla\partial_z T\|_2^2+\alpha\int_t^{t+1}\|\nabla T\|^2_{L^2(\Gamma_u)}+\frac{1}{Re_1}\int_t^{t+1}\|\Delta
v\|^2_2+\frac{1}{Re_2}\int_t^{t+1}\|\nabla\partial_z v\|^2_2\leq \rho_{10}
\end{align}
for any $t\geq T_2+7.$
\subsubsection{$(L^6(\Omega))^2$ estimates of $v_z$}
Multiplying \eqref{3.7.1} by $|u|^4u$ and integrating over $\Omega,$ and combining the boundary
condition \eqref{3.7.2}, we find
\begin{align}\label{3.10.1}
\nonumber&\frac{1}{6}\frac{d}{dt}\|u\|_6^6+\frac{1}{Re_1}\int_{\Omega}|\nabla u|^2|u|^4\,dxdydz+\frac{1}{Re_2}\int_{\Omega}|\partial_z u|^2|u|^4\,dxdydz+\frac{4}{9}\||u|^3\|^2\\
=&\int_{\Omega}(\nabla\cdot v)|u|^6\,dxdydz+\int_{\Omega}\frac{bP}{p}\nabla [(1+aq)T]\cdot |u|^4u\,dxdydz-\int_{\Omega}[(u\cdot\nabla)v]\cdot |u|^4u\,dxdydz.
\end{align}

Next, we estimate each term of the right hand side of \eqref{3.10.1}.
\begin{align}
\nonumber\label{3.10.2}|\int_{\Omega}(\nabla\cdot v)|u|^6\,dxdydz|
\nonumber\leq &C\int_{\Omega}|v||\nabla |u|^3||u|^3\,dxdydz\\
\nonumber\leq&C\|v\|_6\||u|^3\|_3\|\nabla |u|^3\|_2\\
\leq&C\|v\|_6\|u\|_6^{\frac{3}{2}}(\|\nabla|u|^3\|_2+\|\partial_z|u|^3\|_2)^{\frac{3}{2}},
\end{align}
\begin{align}
\nonumber\label{3.10.3}|\int_{\Omega}\frac{bP}{p}\nabla [(1+aq)T]\cdot |u|^4u\,dxdydz|\leq &\||u|^3\|^{\frac{5}{3}}_{\frac{10}{3}}\|\frac{bP}{p}\nabla [(1+aq)T]\|_2\\
\nonumber\leq&C\|\frac{bP}{p}\nabla [(1+aq)T\|_2\||u|^3\|^{\frac{2}{3}}_2\||u|^3\|\\
\leq&C(\|\nabla T\|_2+\|\nabla T\|_3\|q\|_6+\|\nabla q\|_3\|T\|_6)\|u\|^2_6\||u|^3\|
\end{align}
and
\begin{align}
\nonumber\label{3.10.4}|-\int_{\Omega}[(u\cdot\nabla)v]\cdot |u|^4u\,dxdydz|
\leq&C\int_{\Omega}|u|^5|v||\nabla u|\,dxdydz+C\int_{\Omega}|u|^3|v||\nabla |u|^3|\,dxdydz\\
\leq&C\|v\|_6\||u|^3\|_3\|\nabla|u|^3\|_2+C\|v\|_6\||u|^3\|_3\||\nabla u||u|^2\|_2.
\end{align}

  Combining \eqref{3.10.1}-\eqref{3.10.4} with the uniform Gronwall inequality and Young inequality, we have
\begin{align}
\label{3.10.5}\|\partial_z v\|_6^2\leq \rho_{11}
\end{align}
for any $t\geq T_{2}+8.$
\subsubsection{$(L^6(\Omega))^2$ estimates of $(T_z,q_z)$}
Denoted by $\theta=T_z$. It is clear that $\theta$ satisfies the following
 equation by differentiating the equation \eqref{2.5} with respect to
 $z$:
 \begin{align}\label{3.11.1}
\nonumber&\frac{\partial \theta}{\partial t}+L_2\theta+v\cdot\nabla\theta-(\int_0^z\nabla\cdot v(x,y,\zeta,t)d\zeta)\frac{\partial \theta}{\partial
 z}+\partial_{z}v\cdot\nabla T-(\nabla\cdot v)\theta+\frac{bP}{p}(1+aq)(\nabla\cdot v)\\
& -\frac{bP(P-p_0)}{p^2}(1+aq)(\int_0^z\nabla\cdot v(x,y,\zeta,t)d\zeta)+\frac{abP}{p}q_z(\int_0^z\nabla\cdot v(x,y,\zeta,t)d\zeta) =\partial_zQ_1
 \end{align}
supplemented with the boundary conditions
 \begin{align}
\label{3.11.2}(\frac{1}{Rt_2}\theta+\alpha T)|_{\Gamma_u}=0,\theta|_{\Gamma_b}=0,\frac{\partial \theta}{\partial\vec{n}}|_{\Gamma_l}=0.
\end{align}
Taking the inner product of equation
\eqref{3.11.1} with $|\theta|^4\theta$ in $L^2(\Omega)$ and combining the boundary
conditions \eqref{3.11.2}, we know
\begin{align}
\label{3.11.3}\nonumber&\frac{1}{6}\frac{d}{dt}\|\theta\|_6^6+\frac{5}{9Rt_1}\int_{\Omega}|\nabla|\theta|^3|^2\,dxdydz
+\frac{5}{9Rt_2}\int_{\Omega}|\partial_z|\theta|^3|^2\,dxdydz+\alpha^5 Rt_2^4\int_{\Gamma_u}\frac{\partial \theta}{\partial z}|T|^4T\,dxdy\\
\nonumber=&\int_{\Omega}(\nabla\cdot v)|\theta|^6\,dxdydz+\int_{\Omega}\partial_zQ_1|\theta|^4\theta\,dxdydz-\int_{\Omega}(v_z\cdot\nabla T)|\theta|^4\theta\,dxdydz-\int_{\Omega}\frac{bP}{p}(1+aq)(\nabla\cdot v)|\theta|^4\theta\\
&-\int_{\Omega}\frac{abP}{p}q_z(\int_0^z\nabla\cdot v(x,y,\zeta,t)d\zeta)|\theta|^4\theta+\int_{\Omega}\frac{bP(P-p_0)}{p^2}(1+aq)(\int_0^z\nabla\cdot v(x,y,\zeta,t)d\zeta)|\theta|^4\theta.
 \end{align}

 In the following, we give the estimates of each term in the right hand side of
\eqref{3.11.3}.
\begin{align}
\nonumber\label{3.11.4}\left|\int_{\Omega}(\nabla\cdot v)|\theta|^6\,dxdydz\right|\leq &C\int_{\Omega}|v||\nabla |\theta|^3||\theta|^3\,dxdydz\\
\nonumber\leq&C\|v\|_6\||\theta|^3\|_3\|\nabla |\theta|^3\|_2\\
\leq&C\|v\|_6\||\theta|^3\|_2^{\frac{1}{2}}(\|\nabla|\theta|^3\|_2+\|\partial_z|\theta|^3\|_2)^{\frac{3}{2}},
\end{align}
\begin{align}
\nonumber\label{3.11.5}\left|\int_{\Omega}\partial_zQ_1|\theta|^4\theta\,dxdydz\right|\leq &C\int_{\Omega}|\partial_zQ_1||\theta|^5\,dxdydz\\
\nonumber\leq&C\|\partial_zQ_1\|_2\||\theta|^3\|_{\frac{10}{3}}^{\frac{5}{3}}\\
\leq&C\|Q_1\|_{H^1(\Omega)}\||\theta|^3\|_2^{\frac{2}{3}}\||\theta|^3\|_{H^1(\Omega)},
\end{align}
\begin{align}
\nonumber\label{3.11.6}\left|\int_{\Omega}(v_z\cdot\nabla T) |\theta|^4\theta\,dxdydz\right|
\nonumber\leq &C\int_{\Omega}|v_z||\nabla T| |\theta|^5\,dxdydz\\
\nonumber\leq&C\|v_z\|_6\|\nabla T\|_3\||\theta|^3\|_{\frac{10}{3}}^{\frac{5}{3}}\\
\leq&C\|v_z\|_6\|\nabla T\|_2^{\frac{1}{2}}\|\nabla T\|_{H^1(\Omega)}^{\frac{1}{2}}\||\theta|^3\|_2^{\frac{2}{3}}\||\theta|^3\|_{H^1(\Omega)},
\end{align}
\begin{align}
\label{3.11.7}\nonumber\left|\int_{\Omega}\frac{bP}{p}(1+aq)(\nabla\cdot v)|\theta|^4\theta\,dxdydz\right|
\nonumber\leq&C(\|\nabla v\|_2+\|q\|_6\|\nabla v\|_3)\||\theta|^3\|_{\frac{10}{3}}^{\frac{5}{3}}\\
\leq&C(\|\nabla v\|_2+\|q\|_6\|\nabla v\|_3)\||\theta|^3\|_2^{\frac{2}{3}}\||\theta|^3\|_{H^1(\Omega)},
\end{align}
\begin{align}
\label{3.11.8}\nonumber\left|\int_{\Omega}\frac{abP}{p}q_z(\int_0^z\nabla\cdot v(x,y,\zeta,t)d\zeta)|\theta|^4\theta\,dxdydz\right|
\nonumber\leq&C\|q_z\|_6\|\nabla v\|_3\||\theta|^3\|_{\frac{10}{3}}^{\frac{5}{3}}\\
\leq&C\|q_z\|_6\|\nabla v\|_3\||\theta|^3\|_2^{\frac{2}{3}}\||\theta|^3\|_{H^1(\Omega)},\\
\label{3.11.9}\left| \int_{\Omega}\frac{bP(P-p_0)}{p^2}(1+aq)(\int_0^z\nabla\cdot v(x,y,\zeta,t)d\zeta)|\theta|^4\theta\,dxdydz\right|
\leq&C(\|\nabla v\|_2+\|q\|_6\|\nabla v\|_3)\||\theta|^3\|_2^{\frac{2}{3}}\||\theta|^3\|_{H^1(\Omega)},
\end{align}
\begin{align}
\nonumber\label{3.11.10}\alpha^5 Rt_2^4\int_{\Gamma_u}\frac{\partial \theta}{\partial z}|T|^4T\,dxdy
\nonumber=&\alpha^5 Rt_2^5\int_{\Gamma_u}(\frac{\partial T}{\partial t}+v\cdot\nabla T-\frac{1}{Rt_1}\Delta T-Q_1)|T|^4T\,dxdy\\
\nonumber=&\frac{\alpha^5 Rt_2^5}{6}\frac{d}{dt}\int_{\Gamma_u}|T|^6\,dxdy+\alpha^5 Rt_2^5\int_{\Gamma_u}(v\cdot\nabla T)|T|^4T\,dxdy\\
&+\frac{5\alpha^5 Rt_2^5}{9Rt_1}\int_{\Gamma_u}|\nabla |T|^3|^2\,dxdy-\alpha^5 Rt_2^5\int_{\Gamma_u}Q_1|T|^4T\,dxdy,
\end{align}
\begin{align}
\nonumber\label{3.11.11}&\left|\alpha^5 Rt_2^5\int_{\Gamma_u}(v\cdot\nabla T)|T|^4T\,dxdy-\alpha^5 Rt_2^5\int_{\Gamma_u}Q_1|T|^4T\,dxdy\right|\\
\nonumber\leq&C\|v\|_{L^4(\Gamma_u)}\|\nabla |T|^3\|_{L^2(\Gamma_u)}\||T|^3\|_{L^4(\Gamma_u)}+C\|Q_1\|_{L^2(\Gamma_u)}\||T|^3\|_{L^{\frac{10}{3}}(\Gamma_u)}^{\frac{5}{3}}\\
\leq&C\|v\|_{H^1(\Omega)}\|\nabla |T|^3\|_{L^2(\Gamma_u)}^{\frac{3}{2}}\|T\|_{L^6(\Gamma_u)}^{\frac{3}{2}}+C\|Q_1\|_{H^1(\Omega)}\||T|^3\|_{L^2(\Gamma_u)}\||T|^3\|_{H^1(\Gamma_u)}^{\frac{2}{3}}.
\end{align}

Denoted by $\eta=q_z$. It is clear that $\eta$ satisfies the following
 equation by differentiating the equation \eqref{2.6} with respect to
 $z$:
 \begin{align}\label{3.11.12}
\frac{\partial \eta}{\partial t}+L_3\eta+v\cdot\nabla\eta-(\int_0^z\nabla\cdot v(x,y,\zeta,t)d\zeta)\frac{\partial \eta}{\partial
 z}+\partial_zv\cdot\nabla q-(\nabla\cdot v)\eta =\partial_zQ_2
 \end{align}
supplemented with the boundary conditions
 \begin{align}
\label{3.11.13}(\frac{1}{Rt_4}\eta+\beta q)|_{\Gamma_u}=0,\eta|_{\Gamma_b}=0,\frac{\partial \eta}{\partial\vec{n}}|_{\Gamma_l}=0.
\end{align}
Similarly, we have the following inequality
\begin{align}
\nonumber\label{3.11.14}&\frac{d}{dt}(\|\eta\|_6^6+\beta^5 Rt_4^5\|q\|_{L^6(\Gamma_u)}^6)+\frac{2}{Rt_3}\int_{\Omega}|\nabla|\eta|^3|^2\,dxdydz
+\frac{2}{Rt_4}\int_{\Omega}|\partial_z|\eta|^3|^2\,dxdydz+\frac{2\beta^5 Rt_4^5}{Rt_3}\int_{\Gamma_u}|\nabla |q|^3|^2\,d\tilde{x}\\
\leq&C\|v\|_6^4\|\eta\|_6^6+C\|Q_2\|_{H^1(\Omega)}^2\|\eta\|_6^4+C\|v_z\|_6^2\|\nabla q\|_2\|\nabla q\|_{H^1(\Omega)}\|\eta\|_6^4+C\|v\|_{H^1(\Omega)}^4\|q\|_{L^6(\Gamma_u)}^6+C\|Q_2\|_{H^1(\Omega)}^{\frac{3}{2}}\|q\|_{L^6(\Gamma_u)}^{\frac{9}{2}}.
\end{align}
Employing the uniform Gronwall inequality and Young inequality, using \eqref{3.11.1}-\eqref{3.11.14}, yields
\begin{align}\label{3.11.15}
\|q_z\|_6^2+\|q\|_{L^6(\Gamma_u)}^2+\|T_z\|_6^2+\|T\|_{L^6(\Gamma_u)}^2\leq \rho_{12}
\end{align}
for any $t\geq T_2+9.$
\subsubsection{$H$ estimates of $(v_t,T_t, q_t)$ }
 Denoted by $\pi=v_t,$ $\xi=T_t,$ $\chi=q_t.$ It is clear that $\pi,$ $\xi,$ $\chi$ satisfies the following
 equations by differentiating \eqref{2.4}-\eqref{2.6} with respect to
 $t,$ respectively.
\begin{align}\label{3.12.1}
\nonumber&\frac{\partial \pi}{\partial t}+L_1\pi+(v\cdot\nabla)\pi-\left(\int_0^z\nabla\cdot
v(x,y,\zeta,t)d\zeta\right)\frac{\partial \pi}{\partial
z}+(\pi\cdot\nabla)v-\int_0^z\frac{abP}{p(\zeta)}\nabla[\chi(x,y,\zeta,t)T(x,y,\zeta,t)]\,d\zeta+\frac{1}{Ro}f\pi^{\bot}\\
&+\nabla\partial_t\Phi_s(x,y,t)-\left(\int_0^z\nabla\cdot\pi(x,y,\zeta,t)d\zeta\right)\frac{\partial v}{\partial z}-\int_0^z\frac{bP}{p(\zeta)}\nabla[(1+aq(x,y,\zeta,t))\xi(x,y,\zeta,t)]\,d\zeta=0,\\
\label{3.12.2}\nonumber&\frac{\partial \xi}{\partial
t}+L_2\xi+v\cdot\nabla\xi-\left(\int_0^z\nabla\cdot
v(x,y,\zeta,t)d\zeta\right)\frac{\partial \xi}{\partial
z}+\pi\cdot\nabla T+\frac{abP}{p}\chi\left(\int_0^z\nabla\cdot v(x,y,\zeta,t)\,d\zeta\right)\\
&-\left(\int_0^z\nabla\cdot \pi(x,y,\zeta,t)d\zeta\right)\frac{\partial
T}{\partial z}+\frac{bP}{p}(1+aq)\left(\int_0^z\nabla\cdot \pi(x,y,\zeta,t)\,d\zeta\right)=0,\\
\label{3.12.3}&\frac{\partial \chi}{\partial
t}+L_3\chi+v\cdot\nabla\chi-\left(\int_0^z\nabla\cdot
v(x,y,\zeta,t)d\zeta\right)\frac{\partial \chi}{\partial
z}+\pi\cdot\nabla q-\left(\int_0^z\nabla\cdot \pi(x,y,\zeta,t)d\zeta\right)\frac{\partial
q}{\partial z}=0
\end{align}
subject to the boundary conditions
\begin{align}
\label{3.12.4}&\frac{\partial \pi}{\partial z}|_{\Gamma_u}=0,\frac{\partial\pi}{\partial z}|_{\Gamma_b}=0,\pi\cdot\vec{n}|_{\Gamma_l}=0,\frac{\partial \pi}{\partial\vec{n}}\times\vec{n}|_{\Gamma_l}=0,\\
\label{3.12.5}&(\frac{1}{Rt_2}\frac{\partial \xi}{\partial z}+\alpha \xi)|_{\Gamma_u}=0,\frac{\partial \xi}{\partial z}|_{\Gamma_b}=0,\frac{\partial \xi}{\partial\vec{n}}|_{\Gamma_l}=0,\\
\label{3.12.6}&(\frac{1}{Rt_4}\frac{\partial \chi}{\partial z}+\beta \chi)|_{\Gamma_u}=0,\frac{\partial \chi}{\partial z}|_{\Gamma_b}=0,\frac{\partial \chi}{\partial\vec{n}}|_{\Gamma_l}=0.
\end{align}
Multiplying \eqref{3.12.1}, \eqref{3.12.2}, \eqref{3.12.3} by $\pi,$ $\xi,$ $\chi$ respectively, integrating over $\Omega,$ we obtain
\begin{align}\label{3.12.7}
\nonumber\frac{1}{2}\frac{d}{dt}\|\pi\|_2^2+\|\pi\|^2=& -\int_{\Omega}\left[(\pi\cdot\nabla)v-\left(\int_0^z\nabla\cdot
\pi(x,y,\zeta,t)d\zeta\right)\frac{\partial v}{\partial z}\right]\cdot\pi\,dxdydz\\
\nonumber&+\int_{\Omega}\int_0^z\frac{abP}{p(\zeta)}\nabla[\chi(x,y,\zeta,t)T(x,y,\zeta,t)]\,d\zeta\cdot\pi\,dxdydz\\
&+\int_{\Omega}\int_0^z\frac{bP}{p(\zeta)}\nabla[(1+aq(x,y,\zeta,t))\xi(x,y,\zeta,t)]\,d\zeta\cdot\pi\,dxdydz,
\end{align}

\begin{align}\label{3.12.8}
\nonumber\frac{1}{2}\frac{d}{dt}\|\xi\|_2^2+\|\xi\|^2=&-\int_{\Omega}(\pi\cdot\nabla T)\xi\,dxdydz
+\int_{\Omega}\left(\int_0^z\nabla\cdot \pi(x,y,\zeta,t)d\zeta\right)\frac{\partial
T}{\partial z}\xi\,dxdydz\\
\nonumber&-\int_{\Omega}\frac{abP}{p}\chi\left(\int_0^z\nabla\cdot v(x,y,\zeta,t)\,d\zeta\right)\xi\,dxdydz\\
&-\int_{\Omega}\frac{bP}{p}(1+aq)\left(\int_0^z\nabla\cdot \pi(x,y,\zeta,t)\,d\zeta\right)\xi\,dxdydz
\end{align}
and
\begin{align}\label{3.12.9}
\frac{1}{2}\frac{d}{dt}\|\chi\|_2^2+\|\chi\|^2=-\int_{\Omega}(\pi\cdot\nabla q)\chi\,dxdydz
+\int_{\Omega}\left(\int_0^z\nabla\cdot \pi(x,y,\zeta,t)d\zeta\right)\frac{\partial
q}{\partial z}\chi\,dxdydz.
\end{align}

Next, we estimate the right hand side of
\eqref{3.12.7}-\eqref{3.12.9} term by term.
\begin{align}
\label{3.12.10}&\left|-\int_{\Omega}\left[(\pi\cdot\nabla)v-(\int_0^z\nabla\cdot
\pi(x,y,\zeta,t)d\zeta)\frac{\partial v}{\partial z}\right]\cdot\pi\,dxdydz\right|
\leq C\|\nabla\pi\|_2\|v_z\|_6\|\pi\|_3+C\|v\|_6\|\pi\|_3\|\nabla\pi\|_2,\\
\label{3.12.11}\nonumber&\left|-\int_{\Omega}(\pi\cdot\nabla T)\xi\,dxdydz
+\int_{\Omega}\left(\int_0^z\nabla\cdot \pi(x,y,\zeta,t)d\zeta\right)\frac{\partial
T}{\partial z}\xi\,dxdydz\right|\\
\leq&C\|\nabla\pi\|_2\|T\|_6\|\xi\|_3+C\|\nabla\xi\|_2\|T\|_6\|\pi\|_3+C\|\nabla\pi\|_2\|T_z\|_6\|\xi\|_3,
\end{align}
\begin{align}
\label{3.12.12}\nonumber&\left|-\int_{\Omega}(\pi\cdot\nabla q)\chi\,dxdydz
+\int_{\Omega}\left(\int_0^z\nabla\cdot \pi(x,y,\zeta,t)d\zeta\right)\frac{\partial
q}{\partial z}\chi\,dxdydz\right|\\
\leq&C\|\nabla\pi\|_2\|q\|_6\|\chi\|_3+C\|\nabla\chi\|_2\|q\|_6\|\pi\|_3+C\|\nabla\pi\|_2\|q_z\|_6\|\chi\|_3,\\
\label{3.12.13}&\left|\int_{\Omega}\int_0^z\frac{abP}{p(\zeta)}\nabla[\chi(x,y,\zeta,t)T(x,y,\zeta,t)]\,d\zeta\cdot\pi\,dxdydz\right|
\leq C\|\chi\|_3\|T\|_6\|\nabla\pi\|_2,\\
\label{3.12.14}&\left|-\int_{\Omega}\frac{abP}{p}\chi\left(\int_0^z\nabla\cdot v(x,y,\zeta,t)\,d\zeta\right)\xi\,dxdydz\right|
\leq C\|\nabla\chi\|_2\|v\|_6\|\xi\|_3+C\|\nabla\xi\|_2\|v\|_6\|\chi\|_3
\end{align}
and
\begin{align}\label{3.12.15}
\nonumber&\int_{\Omega}\int_0^z\frac{bP}{p(\zeta)}\nabla[(1+aq(x,y,\zeta,t))\xi(x,y,\zeta,t)]\,d\zeta\cdot\pi\,dxdydz\\
&-\int_{\Omega}\frac{bP}{p}(1+aq)\left(\int_0^z\nabla\cdot \pi(x,y,\zeta,t)\,d\zeta\right)\xi\,dxdydz=0.
\end{align}
We infer from \eqref{3.12.7}-\eqref{3.12.15} that
\begin{align}\label{3.12.16}
\nonumber&\frac{d}{dt}(\|\xi\|_2^2+\|\pi\|_2^2+\|\chi\|_2^2)+(\|\xi\|^2+\|\pi\|^2+\|\chi\|^2)\\
\leq &(\|v\|_6^4+\|v_z\|_6^4+\|T\|_6^4+\|T_z\|_6^4+\|q\|_6^4+\|q_z\|_6^4)(\|\xi\|_2^2+\|\pi\|_2^2+\|\chi\|_2^2).
\end{align}
Taking the inner product of \eqref{2.6} with $\chi$ in $L^2(\Omega),$ we obtain
\begin{align}\label{3.12.17}
\nonumber\|\chi\|_2^2=&-\int_{\Omega}\left[v\cdot\nabla q-\left(\int_0^z
\nabla\cdot v(x,y,\zeta,t)\,d\zeta\right)\frac{\partial q}{\partial
z}\right]\chi
\,dxdydz-\int_{\Omega}L_3 q\chi\,dxdydz+\int_{\Omega}Q_2\chi\, dxdydz\\
\leq&\|Q_2\|_2\|\chi\|_2+C\|v\|_6\|\nabla
q\|_3\|\chi\|_2+\|L_3q\|_2\|\chi\|_2+C\|\nabla v\|_3\|q_z\|_6\|\chi\|_2.
\end{align}
Similarly, we have
\begin{align}\label{3.12.18}
\|\xi\|_2^2\leq \|Q_1\|_2\|\xi\|_2+C\|v\|_6\|\nabla T\|_3\|\xi\|_2+\|\nabla v\|_3\|T_z\|_6\|\xi\|_2+\|L_2T\|_2\|\xi\|_2+C\|\nabla v\|_2\|\xi\|_2+C\|q\|_6\|\nabla v\|_3\|\xi\|_2
\end{align}
and
\begin{align}\label{3.12.19}
\nonumber\|\pi\|_2^2\leq &C\|v\|_6\|\nabla v\|_3\|\pi\|_2+C\|\nabla v\|_3\|v_z\|_6\|\pi\|_2+\|L_1v\|_2\|\pi\|_2+C\|v\|_2\|\pi\|_2\\
&+C\|\nabla T\|_2\|\pi\|_2+C\|q\|_6\|\nabla T\|_3\|\pi\|_2+C\|T\|_6\|\nabla q\|_3\|\pi\|_2.
\end{align}
By the uniform Gronwall inequality and Young inequality, we derive from \eqref{3.12.16}-\eqref{3.12.19} that
\begin{align}\label{3.12.10}
\|v_t\|_2^2+\|T_t\|_2^2+\|q_t\|_2^2\leq \rho_{13}
\end{align}
for any $t\geq T_2+10.$

Moreover, we have
\begin{align}\label{3.12.21}
\int_t^{t+1}\|v_t\|^2\,d\tau+\int_t^{t+1}\|T_t\|^2\,d\tau+\int_t^{t+1}\|q_t\|^2\,d\tau\leq\rho_{14}
\end{align}
for any $t\geq T_2+10.$
\subsubsection{$V$ estimates of $(v_t,T_t, q_t)$ }
Multiplying \eqref{3.12.1} by $L_1\pi$ and integrating over $\Omega,$ we get
\begin{align*}
\frac{1}{2}\frac{d}{dt}\|\pi\|^2+\|L_1\pi\|_2^2
\leq&C\|v\|_6\|\nabla\pi\|_3\|L_1\pi\|_2+C\|\nabla v\|_2^{\frac{1}{2}}\|\Delta v\|_2^{\frac{1}{2}}\|\pi_z\|_2^{\frac{1}{2}}\|\nabla \pi_z\|_2^{\frac{1}{2}}\|L_1\pi\|_2+C\|\chi\|_6\|\nabla T\|_3\|L_1\pi\|_2\\
&+C\|\pi\|_6\|\nabla v\|_3\|L_1\pi\|_2+C\|\nabla\pi\|_2^{\frac{1}{2}}\|\Delta \pi\|_2^{\frac{1}{2}}\|v_z\|_2^{\frac{1}{2}}\|\nabla v_z\|_2^{\frac{1}{2}}\|L_1\pi\|_2+C\|\nabla\chi\|_3\|T\|_6\|L_1\pi\|_2\\
&+C\|\nabla\xi\|_2\|L_1\pi\|_2+C\|q\|_6\|\nabla\xi\|_3\|L_1\pi\|_2+C\|\xi\|_6\|\nabla q\|_3\|L_1\pi\|_2.
\end{align*}
Similarly, we have the following inequalities
\begin{align*}
\frac{1}{2}\frac{d}{dt}\|\xi\|^2+\|L_2\xi\|_2^2
\leq&C\|v\|_6\|\nabla\xi\|_3\|L_2\xi\|_2+C\|\nabla v\|_2^{\frac{1}{2}}\|\Delta v\|_2^{\frac{1}{2}}\|\xi_z\|_2^{\frac{1}{2}}\|\nabla \xi_z\|_2^{\frac{1}{2}}\|L_2\xi\|_2+C\|\nabla \pi\|_2\|L_2\xi\|_2\\
&+C\|\pi\|_6\|\nabla T\|_3\|L_2\xi\|_2+C\|\nabla\pi\|_2^{\frac{1}{2}}\|\Delta \pi\|_2^{\frac{1}{2}}\|T_z\|_2^{\frac{1}{2}}\|\nabla T_z\|_2^{\frac{1}{2}}\|L_2\xi\|_2+C\|\nabla v\|_3\|\chi\|_6\|L_2\xi\|_2\\
&+C\|q\|_6\|\nabla\pi\|_3\|L_2\xi\|_2
\end{align*}
and
\begin{align*}
\frac{1}{2}\frac{d}{dt}\|\chi\|^2+\|L_3\chi\|_2^2\leq&C\|v\|_6\|\nabla\chi\|_3\|L_3\chi\|_2+C\|\nabla v\|_2^{\frac{1}{2}}\|\Delta v\|_2^{\frac{1}{2}}\|\chi_z\|_2^{\frac{1}{2}}\|\nabla \chi_z\|_2^{\frac{1}{2}}\|L_3\chi\|_2\\
&+C\|\pi\|_6\|\nabla q\|_3\|L_3\chi\|_2+C\|\nabla \pi\|_2^{\frac{1}{2}}\|\Delta\pi\|_2^{\frac{1}{2}}\|q_z\|_2^{\frac{1}{2}}\|\nabla q_z\|_2^{\frac{1}{2}}\|L_3\chi\|_2.
\end{align*}
Employing the uniform Gronwall inequality and Young inequality, yield
\begin{align}\label{3.13.1}
\|v_t\|^2+\|T_t\|^2+\|q_t\|^2\leq \rho_{15}
\end{align}
for any $t\geq T_2+11.$
\subsubsection{$(H^2(\Omega))^4\cap V$ estimates of $(v,T,q)$}
Taking the inner product of \eqref{2.4} with $L_1v$ in $L^2(\Omega),$ we obtain
\begin{align*}
\|L_1v\|_2^2\leq&C\|v\|_6\|\nabla v\|_3\|L_1v\|_2+C\|\nabla v\|_3\|v_z\|_6\|L_1v\|_2+C\|q\|_6\|\nabla T\|_3\|L_1v\|_2\\
&+C\|v\|_2\|L_1v\|_2+C\|\nabla T\|_2\|L_1v\|_2+\|v_t\|_2\|L_1v\|_2+\|\nabla q\|_3\|T\|_6\|L_1v\|_2.
\end{align*}
Similarly, we have the following inequalities
\begin{align*}
\|L_2T\|_2^2\leq&C\|v\|_6\|\nabla T\|_3\|L_2T\|_2+C\|\nabla v\|_3\|T_z\|_6\|L_2T\|_2+C\|q\|_6\|\nabla v\|_3\|L_2T\|_2\\
&+C\|Q_1\|_2\|L_2T\|_2+C\|\nabla v\|_2\|L_2T\|_2+\|T_t\|_2\|L_2T\|_2
\end{align*}
and
\begin{align*}
\|L_3q\|_2^2\leq C\|v\|_6\|\nabla q\|_3\|L_3q\|_2+C\|\nabla v\|_3\|q_z\|_6\|L_3q\|_2+C\|Q_2\|_2\|L_3q\|_2+\|q_t\|_2\|L_3q\|_2.
\end{align*}
Therefore, we have
\begin{align*}
\|L_1v\|_2^2+\|L_2T\|_2^2+\|L_3q\|_2^2\leq\rho_{16}
\end{align*}
for any $t\geq T_2+10,$ which implies that
\begin{align}\label{3.14.1}
\|(v,T,q)\|_{(H^2(\Omega))^4\cap V}^2\leq\rho_{17}
\end{align}
for any $t\geq T_2+10.$

By virtue of \eqref{3.1.3}, \eqref{3.2.9}, \eqref{3.7.8}, \eqref{3.8.5}, \eqref{3.9.7}, \eqref{3.14.1}, we have
\begin{theorem}\label{3.2}
Assume that $Q_1 \in L^2(\Omega)$ and $Q_2 \in L^2(\Omega).$ Then the semigroup $\{S(t)\}_{t \geq
0}$ associated with the initial-boundary problem \eqref{2.4}-\eqref{2.12} possesses an absorbing set in $V.$
That is, there exists a positive constant $\mathcal{R}_1$ satisfying for any bounded subset $B$ of $V,$ there exists a positive time $\tau_1=\tau_{1,B}$
 depending on the norm of $B$ such that for any $t \geq \tau_1,$ we have
\begin{align*}
\|(v(t),T(t),q(t))\|_V=& \|S(t)(v_0,T_0,q_0)\|_V\leq\mathcal{R}_1.
\end{align*}
\end{theorem}
\begin{theorem}\label{3.3}
Assume that $Q_1 \in H^1(\Omega)$ and $Q_2 \in H^1(\Omega).$ Then the semigroup $\{S(t)\}_{t \geq
0}$ associated with the initial-boundary problem \eqref{2.4}-\eqref{2.12} possesses an absorbing set in $(H^2(\Omega))^4\cap V.$
That is, there exists a positive constant $\mathcal{R}_2$ satisfying for any bounded subset $B$ of $V,$ there exists a positive time $\tau_2=\tau_{2,B}$
 depending on the norm of $B$ such that for any $t \geq \tau_2,$ we have
\begin{align*}
\|(v(t),T(t),q(t))\|_{(H^2(\Omega))^4\cap V}=& \|S(t)(v_0,T_0,q_0)\|_{(H^2(\Omega))^4\cap V}\leq \mathcal{R}_2.
\end{align*}
\end{theorem}
\section{The existence of attractors}
\def\theequation{4.\arabic{equation}}\makeatother
\setcounter{equation}{0}
\subsection{The existence of global attractors}
 The abstract theory of global attractor can be referred to \cite{bav, cvv, mqf, tr, ymh, zck}. In this section, we prove the existence of global attractors of the semigroup $\{S(t)\}_{t \geq0}$ generated by the initial-boundary problem \eqref{2.4}-\eqref{2.12}.

Thanks to the compactness of $(H^2(\Omega))^4\cap V\subset V,$ we have the following result.
\begin{theorem}\label{4.1.0}
Assume that $Q_1\in H^1(\Omega)$ and $Q_2\in H^1(\Omega).$ Then the semigroup $\{S(t)\}_{t \geq 0}$
corresponding to problem \eqref{2.4}-\eqref{2.12} has a global
attractor $\mathcal{A}_V$ in $V.$
\end{theorem}

\begin{remark}
Under the assumptions that $Q_1\in L^2(\Omega)$ and $Q_2\in L^2(\Omega),$ the existence of a global
attractor $\mathcal{A}_V$ in $V$ of the semigroup $\{S(t)\}_{t \geq 0}$
associated with problem \eqref{2.4}-\eqref{2.12} can be also obtained by the Aubin-Lions compactness Lemma as in \cite{jn}.
\end{remark}

Next, we prove the asymptotical compactness of the semigroup $\{S(t)\}_{t \geq 0}$
generated by the initial-boundary problem \eqref{2.4}-\eqref{2.12}.

\begin{theorem}\label{4.1.1}
Assume that $Q_1\in H^1(\Omega)$ and $Q_2\in H^1(\Omega).$ Then the semigroup $\{S(t)\}_{t \geq 0}$
generated by problem \eqref{2.4}-\eqref{2.12} is asymptotically compact in $(H^2(\Omega))^4\cap V.$
\end{theorem}
\textbf{Proof.}
Let $B_0$ be an absorbing set in $(H^2(\Omega))^4\cap V$ of the semigroup $\{S(t)\}_{t \geq 0}$ generated by problem \eqref{2.4}-\eqref{2.12} obtained in Theorem \ref{3.3}.
 Then we need only to show that
 for any $\{(v_{0n},T_{0n},q_{0n})\}_{n=1}^{\infty}\subset B_0$ and $t_n\rightarrow
 \infty$, $\{(v_n(t_n),T_n(t_n),q_n(t_n))\}_{n=1}^{\infty}$ is pre-compact in $(H^2(\Omega))^4\cap V,$ where $(v_n(t_n),T_n(t_n),q_n(t_n))=S(t_n)(v_{0n},T_{0n},q_{0n}).$

In fact, from Theorem \ref{3.3}, \eqref{3.13.1} and the compactness of $(H^2(\Omega))^4\cap V\subset (W^{1,3}(\Omega))^4\cap V,$ we know that
$\{(v_n(t_n), T_n(t_n), q_n(t_n))\}_{n=1}^{\infty}$ and $\{(\frac{\partial v_n(t_n)}{\partial t}, \frac{\partial T_n(t_n)}{\partial t}, \frac{\partial q_n(t_n)}{\partial t})\}_{n=1}^{\infty}$ is
pre-compact in $(W^{1,3}(\Omega))^4\cap V$ and $H,$ respectively. Without loss of
generality, we assume that
$\{(v_n(t_n), T_n(t_n), q_n(t_n))\}_{n=1}^{\infty}$ and $\{(\frac{\partial v_n(t_n)}{\partial t}, \frac{\partial T_n(t_n)}{\partial t}, \frac{\partial q_n(t_n)}{\partial t})\}_{n=1}^{\infty}$ is a Cauchy
sequence in $(W^{1,3}(\Omega))^4\cap V$ and $H,$ respectively.

 In the following, we will prove
that $\{(v_n(t_n), T_n(t_n), q_n(t_n))\}_{n=0}^{\infty}$ is a Cauchy sequence in $(H^2(\Omega))^4\cap V.$

Then, by simply calculations, we have
\begin{align*}
\|L_3q_n(t_n)-L_3q_m(t_m)\|_2^2\leq&\|\frac{\partial q_n(t_n)}{\partial t}-\frac{\partial q_m(t_m)}{\partial t}\|_2\|L_3q_n(t_n)-L_3q_m(t_m)\|_2+\|v_n(t_n)-v_m(t_m)\|_3\|\nabla q_n(t_n)\|_6\|L_3q_n(t_n)-L_3q_m(t_m)\|_2\\
&+\|v_m(t_m)\|_6\|\nabla q_n(t_n)-\nabla q_m(t_m)\|_3\|L_3q_n(t_n)-L_3q_m(t_m)\|_2\\
&+\|\nabla v_n(t_n)-\nabla v_m(t_m)\|_3\|\frac{\partial q_n(t_n)}{\partial z}\|_6\|L_3q_n(t_n)-L_3q_m(t_m)\|_2\\
&+\|\nabla v_m(t_m)\|_6\|\frac{\partial q_n(t_n)}{\partial z}-\frac{\partial q_m(t_m)}{\partial z}\|_3\|L_3q_n(t_n)-L_3q_m(t_m)\|_2.
\end{align*}
From H\"{o}lder inequality and Theorem \ref{3.3}, we deduce that
$\{q_n(t_n)\}_{n=1}^{\infty}$ is a Cauchy sequence in $H^2(\Omega)$.

Similarly, we can also prove $\{(v_n(t_n),T_n(t_n))\}_{n=1}^{\infty}$ is a Cauchy sequence in $(H^2(\Omega))^3$.
The proof of Theorem \ref{4.1.1} is completed.\\
\qed\hfill

Therefore, from Theorem \ref{3.3} and Theorem \ref{4.1.1},
we immediately obtain the following result.
\begin{theorem}
Assume that $Q_1\in H^1(\Omega)$ and $Q_2\in H^1(\Omega).$ Then the semigroup $\{S(t)\}_{t \geq 0}$
associated with problem \eqref{2.4}-\eqref{2.12} has a global
attractor $\mathcal{A}$ in $(H^2(\Omega))^4\cap V.$
\end{theorem}
\subsection{The existence of an exponential attractor}
In this section, inspired by the idea in \cite{em}, we prove the existence of an exponential attractor in $V$ for the three dimensional viscous primitive equations of large-scale moist atmosphere. The definition about exponential attractor can be referred to \cite{cvv, em, em1}.

In order to estimate the fractal dimension of the exponential attractor in $V,$ we need the following lemma.
\begin{lemma}(\cite{mj})\label{4.2.0}
Let $X$ and $Y$ be two metric spaces and $f:X\rightarrow Y$ be $\alpha$-H\"{o}lder continuous on the subset $A\subset X.$ Then
\begin{align*}
d_F(f(A),Y)\leq\frac{1}{\alpha}d_F(A,X).
\end{align*}
In particular, the fractal dimension does not increase under a Lipschitz continuous mapping.
\end{lemma}
In the following, we first prove the first Theorem about the smoothing property of the semigroup $\{S(t)\}_{t\geq0}$ generated by problem \eqref{2.4}-\eqref{2.12}.
\begin{theorem}\label{4.2.2}
Assume that $Q_1\in H^1(\Omega)$ and $Q_2\in H^1(\Omega).$ Let $(v^i,\Phi^i_s,T^i,q^i)$ be the solution of problem \eqref{2.4}-\eqref{2.12} with the initial data $(v_0^i,T_0^i,q_0^i)\in V,$ $i=1,2.$ Then the following estimate holds
\begin{align}
\|(v^1(t),T^1(t),q^1(t))-(v^2(t),T^2(t),q^2(t))\|_V^2\leq \varrho_1\frac{\bar{t}+1}{\bar{t}}e^{\varrho_2 t}\|(v_0^1,T_0^1,q_0^1)-(v_0^2,T_0^2,q_0^2)\|_H^2
\end{align}
for any $t\geq\tau_2,$ where $\bar{t}=t-\tau_2,$ $\varrho_1$ and $\varrho_2$ are positive constants which only depend on $\Omega,$ $\alpha,$ $\beta,$ $Rt_2,$ $Rt_4,$ $\|Q_1\|_{H^1(\Omega)}$ and $\|Q_2\|_{H^1(\Omega)}.$
\end{theorem}
\textbf{Proof.} Let $(v,\Phi_s,T,q)=(v^1-v^2,\Phi_s^1-\Phi_s^2,T^1-T^2,q^1-q^2),$ then $(v,\Phi_s,T,q)$ satisfies the following equations
\begin{equation}\label{4.2.3}
\begin{cases}
&\frac{\partial v}{\partial t}+(v^1\cdot
\nabla)v-(\int_0^z\nabla\cdot
v^1(x,y,\zeta,t)\,d\zeta)\frac{\partial v}{\partial z}+(v\cdot
\nabla)v^2-(\int_0^z\nabla\cdot
v(x,y,\zeta,t)\,d\zeta)\frac{\partial v^2}{\partial z}+\nabla
\Phi_s(x,y,t)+\frac{1}{Ro}fv^{\bot}\\
&+L_1v-\int_0^{z}\frac{bP}{p(\zeta)}\nabla [(1+aq^1(x,y,\zeta,t))T(x,y,\zeta,t)]\,d\zeta-\int_0^{z}\frac{abP}{p(\zeta)}\nabla [q(x,y,\zeta,t)T^2(x,y,\zeta,t)]\,d\zeta=0,\\
&\frac{\partial T}{\partial t}+v^1\cdot \nabla
T-(\int_0^z\nabla\cdot v^1(x,y,\zeta,t)\,d\zeta)\frac{\partial T}{\partial z}+v\cdot \nabla
T^2-(\int_0^z\nabla\cdot v(x,y,\zeta,t)\,d\zeta)\frac{\partial T^2}{\partial z}+L_2T\\
&+\frac{bP}{p}(1+aq^1)(\int_0^z\nabla\cdot v(x,y,\zeta,t)\,d\zeta)+\frac{abP}{p}q(\int_0^z\nabla\cdot v^2(x,y,\zeta,t)\,d\zeta)=0,\\
& \frac{\partial q}{\partial t}+v^1\cdot\nabla
q-(\int_0^z\nabla\cdot v^1(x,y,\zeta,t)\,d\zeta)\frac{\partial
q}{\partial z}+v\cdot\nabla
q^2-(\int_0^z\nabla\cdot v(x,y,\zeta,t)\,d\zeta)\frac{\partial
q^2}{\partial z}+L_3q=0
\end{cases}
\end{equation}
with the following boundary conditions
\begin{equation}\label{4.2.4}
\begin{cases}
&\frac{\partial v}{\partial z}|_{\Gamma_u}=0,\frac{\partial
v}{\partial z}|_{\Gamma_b}=0,v\cdot
\vec{n}|_{\Gamma_l}=0,\frac{\partial v}{\partial
\vec{n}}\times\vec{
n}|_{\Gamma_l}=0,\\
&(\frac{1}{Rt_2}\frac{\partial T}{\partial z}+\alpha
T)|_{\Gamma_u}=0,\frac{\partial T}{\partial z}|_{\Gamma_b}=0,
\frac{\partial T}{\partial\vec{n}}|_{\Gamma_l}=0,\\
&(\frac{1}{Rt_4}\frac{\partial q}{\partial z}+\beta
q)|_{\Gamma_u}=0,\frac{\partial q}{\partial z}|_{\Gamma_b}=0,
\frac{\partial q}{\partial\vec{n}}|_{\Gamma_l}=0
\end{cases}
\end{equation}
and the initial data
\begin{equation}\label{4.2.5}
\begin{cases}
&v(x,y,z,0)=v_0^1(x,y,z)-v_0^2(x,y,z),\\
&T(x,y,z,0)=T_0^1(x,y,z)-T_0^2(x,y,z),\\
&q(x,y,z,0)=q_0^1(x,y,z)-q_0^2(x,y,z).
\end{cases}
\end{equation}
Multiplying the first equation of \eqref{4.2.3} by $v$ and integrating over $\Omega,$ we obtain
\begin{align}\label{4.2.6}
\nonumber&\frac{1}{2}\frac{d}{dt}\|v\|_2^2+\|v\|^2=\int_\Omega [(v\cdot
\nabla)v^2]\cdot v\,dxdydz-\int_\Omega\int_0^{z}\frac{bP}{p(\zeta)}\nabla [(1+aq^1(x,y,\zeta,t))T(x,y,\zeta,t)]\,d\zeta\cdot v\,dxdydz\\
&-\int_\Omega(\int_0^z\nabla\cdot
v(x,y,\zeta,t)\,d\zeta)\frac{\partial v^2}{\partial z}\cdot v\,dxdydz-\int_\Omega\int_0^{z}\frac{abP}{p(\zeta)}\nabla [q(x,y,\zeta,t)T^2(x,y,\zeta,t)]\,d\zeta\cdot v\,dxdydz.
\end{align}
Taking the inner product of the second equation of \eqref{4.2.3} with $T$ in $L^2(\Omega)$ and combining the second equation of \eqref{4.2.4}, we get
\begin{align}\label{4.2.7}
\nonumber&\frac{1}{2}\frac{d}{dt}\|T\|_2^2+\|T\|^2=\int_\Omega (v\cdot \nabla
T^2)T\,dxdydz+\int_\Omega\frac{bP}{p}(1+aq^1)(\int_0^z\nabla\cdot v(x,y,\zeta,t)\,d\zeta)T\,dxdydz\\
&-\int_\Omega(\int_0^z\nabla\cdot v(x,y,\zeta,t)\,d\zeta)\frac{\partial T^2}{\partial z}T\,dxdydz+\int_\Omega\frac{abP}{p}q(\int_0^z\nabla\cdot v^2(x,y,\zeta,t)\,d\zeta)T\,dxdydz.
\end{align}
Multiplying the third equation of \eqref{4.2.3} by $q$ and integrating over $\Omega,$ we obtain
\begin{align}\label{4.2.8}
\frac{1}{2}\frac{d}{dt}\|q\|_2^2+\|q\|^2
=\int_\Omega(v\cdot\nabla
q^2)q\,dxdydz-\int_\Omega(\int_0^z\nabla\cdot v(x,y,\zeta,t)\,d\zeta)\frac{\partial
q^2}{\partial z}q\,dxdydz.
\end{align}
From \eqref{4.2.6}-\eqref{4.2.8} and H\"{o}lder inequality, we deduce that
\begin{align*}
\frac{d}{dt}\|(v,T,q)\|_H^2+2\|(v,T,q)\|_V^2\leq&C\|v\|_3\|\nabla v^2\|_2\|v\|_6+C\|v_z^2\|_2^{\frac{1}{2}}\|\nabla v_z^2\|_2^{\frac{1}{2}}\|v\|_2^{\frac{1}{2}}\|\nabla v\|_2^{\frac{3}{2}}+C\|q\|_3\|\nabla v\|_2\|T^2\|_6+C\|v\|_3\|\nabla T^2\|_2\|T\|_6\\
&+C\|T_z^2\|_2^{\frac{1}{2}}\|\nabla T_z^2\|_2^{\frac{1}{2}}\|T\|_2^{\frac{1}{2}}\|\nabla T\|_2^{\frac{1}{2}}\|\nabla v\|_2+C\|q\|_6\|\nabla v^2\|_2\|T\|_3+C\|v\|_3\|\nabla q^2\|_2\|q\|_6\\
&+C\|q_z^2\|_2^{\frac{1}{2}}\|\nabla q_z^2\|_2^{\frac{1}{2}}\|q\|_2^{\frac{1}{2}}\|\nabla q\|_2^{\frac{1}{2}}\|\nabla v\|_2.
\end{align*}
It follows from Theorem \ref{3.3} and Young inequality that
\begin{align*}
\frac{d}{dt}\|(v,T,q)\|_H^2+\|(v,T,q)\|_V^2\leq&C\mathcal{R}_2^4\|(v,T,q)\|_H^2
\end{align*}
for any $t\geq \tau_2.$

We infer from the classical Gronwall inequality that
\begin{align*}
\|(v(t),T(t),q(t))\|_H^2+\int_0^t\|(v(s),T(s),q(s))\|_V^2\,ds\leq e^{C\mathcal{R}_2^4t}\|(v(0),T(0),q(0))\|_H^2
\end{align*}
for any $t\geq \tau_2,$ where $C$ is a positive constant.

Taking the inner product of the first equation of \eqref{4.2.3} with $L_1v$ in $L^2(\Omega)$ and combining the first equation of \eqref{4.2.4}, we get
\begin{align}\label{4.2.9}
\nonumber&\frac{1}{2}\frac{d}{dt}\|v\|^2+\|L_1v\|_2^2\leq\|v^1\|_6\|\nabla v\|_3\|L_1v\|_2+C\|\nabla v^1\|_2^{\frac{1}{2}}\|\Delta v^1\|_2^{\frac{1}{2}}\| v_z\|_2^{\frac{1}{2}}\|\nabla v_z\|_2^{\frac{1}{2}}\|L_1v\|_2+\|v\|_3\|\nabla v^2\|_6\|L_1v\|_2\\
\nonumber&+C\|\nabla v\|_2^{\frac{1}{2}}\|\Delta v\|_2^{\frac{1}{2}}\|v^2_z\|_2^{\frac{1}{2}}\|\nabla v^2_z\|_2^{\frac{1}{2}}\|L_1v\|_2+C\|v\|_2\|L_1v\|_2+C\|\nabla T\|_2\|L_1v\|_2+C\|\nabla q^1\|_3\|T\|_6\|L_1v\|_2\\
&+C\|q^1\|_6\|\nabla T\|_3\|L_1v\|_2+C\|\nabla q\|_3\|T^2\|_6\|L_1v\|_2+C\|q\|_6\|\nabla T^2\|_3\|L_1v\|_2.
\end{align}
Multiplying the third equation of \eqref{4.2.3} by $L_2T$ and integrating over $\Omega,$ we obtain
\begin{align}\label{4.2.10}
\nonumber&\frac{1}{2}\frac{d}{dt}\|T\|^2+\|L_2T\|_2^2\leq\|v^1\|_6\|\nabla T\|_3\|L_2T\|_2+C\|\nabla v^1\|_2^{\frac{1}{2}}\|\Delta v^1\|_2^{\frac{1}{2}}\| T_z\|_2^{\frac{1}{2}}\|\nabla T_z\|_2^{\frac{1}{2}}\|L_2T\|_2+\|v\|_3\|\nabla T^2\|_6\|L_2T\|_2\\
&+C\|\nabla v\|_2^{\frac{1}{2}}\|\Delta v\|_2^{\frac{1}{2}}\|T^2_z\|_2^{\frac{1}{2}}\|\nabla T^2_z\|_2^{\frac{1}{2}}\|L_2T\|_2+C\|\nabla v\|_2\|L_2T\|_2+C\|q^1\|_6\|\nabla v\|_3\|L_2T\|_2+C\|q\|_6\|\nabla v^2\|_3\|L_2T\|_2.
\end{align}
Taking the inner product of the third equation of \eqref{4.2.3} with $L_3q$ in $L^2(\Omega)$ and combining the third equation of \eqref{4.2.4}, we get
\begin{align}\label{4.2.11}
\nonumber&\frac{1}{2}\frac{d}{dt}\|q\|^2+\|L_3q\|_2^2\leq\|v^1\|_6\|\nabla q\|_3\|L_3q\|_2+C\|\nabla v^1\|_2^{\frac{1}{2}}\|\Delta v^1\|_2^{\frac{1}{2}}\| q_z\|_2^{\frac{1}{2}}\|\nabla q_z\|_2^{\frac{1}{2}}\|L_3q\|_2+\|v\|_3\|\nabla q^2\|_6\|L_3q\|_2\\
&+C\|\nabla v\|_2^{\frac{1}{2}}\|\Delta v\|_2^{\frac{1}{2}}\|q^2_z\|_2^{\frac{1}{2}}\|\nabla q^2_z\|_2^{\frac{1}{2}}\|L_3q\|_2.
\end{align}
We deduce from \eqref{4.2.9}-\eqref{4.2.11}, Theorem \ref{3.3} and Young inequality that
\begin{align*}
\frac{d}{dt}\|(v,T,q)\|_V^2+\|(L_1v,L_2T,L_3q)\|_H^2\leq C(1+\mathcal{R}_2^4)\|(v,T,q)\|_V^2
\end{align*}
for any $t\geq \tau_2.$

Multiplying now both side of the above inequality by $\bar{t}=t-\tau_2$ and integrating the resulting relation over $(\tau_2,t),$ we obtain
\begin{align*}
\bar{t}\|(v(t),T(t),q(t))\|_V^2\leq &C(1+\mathcal{R}_2^4)\int_{\tau_2}^t(s-\tau_2+1)\|(v(s),T(s),q(s))\|_V^2\,ds\\
\leq&C(1+\mathcal{R}_2^4)(\bar{t}+1)\int_0^t\|(v(s),T(s),q(s))\|_V^2\,ds\\
\leq&C(1+\mathcal{R}_2^4)(\bar{t}+1)e^{C\mathcal{R}_2^4t}\|(v(0),T(0),q(0))\|_H^2
\end{align*}
for any $t\geq \tau_2.$\\
\qed\hfill

The second Theorem is concerned with the time regularity of the semigroup $\{S(t)\}_{t\geq0}$ generated by problem \eqref{2.4}-\eqref{2.12}. The proof is standard and we only state it here.
\begin{theorem}\label{4.2.12}
Assume that $Q_1\in H^1(\Omega)$ and $Q_2\in H^1(\Omega).$ Then for any bounded subset $B\subset V,$ there exists a positive constant $\varrho_3$ and a time $t^*=t^*(B)>0$ such that
\begin{align}
\|S(t)(v_0,T_0,q_0)-S(\tilde{t})(v_0,T_0,q_0)\|_V\leq \varrho_3\|t-\tilde{t}\|
\end{align}
for any $t,\tilde{t}\geq t^*$ and any $(v_0,T_0,q_0)\in B,$ where $S(t)(v_0,T_0,q_0)$ is the solution of problem \eqref{2.4}-\eqref{2.12} with initial data $(v_0,T_0,q_0).$
\end{theorem}

Finally, we prove the existence of an exponential attractor for problem \eqref{2.4}-\eqref{2.12}.
\begin{theorem}\label{4.2.13}
Assume that $Q_1\in H^1(\Omega)$ and $Q_2\in H^1(\Omega).$ Let $\{S(t)\}_{t\geq
0}$ be a semigroup generated by problem \eqref{2.4}-\eqref{2.12}. Then the semigroup $\{S(t)\}_{t\geq
0}$ possesses an exponential attractor $\mathcal{E}\subset V,$ namely,
\begin{itemize}
\item [(i)] $\mathcal{E}$ is compact and positively invariant with respect to $S(t),$ i.e.,
\begin{align*}
S(t)\mathcal{E}\subset\mathcal{E}
\end{align*}
for any $t\geq0.$
\end{itemize}
\item [(ii)] The fractal dimension $dim_F(\mathcal{E},V)$ of $\mathcal{E}$ is finite.
\item [(iii)] $\mathcal{E}$ attracts exponentially any bounded subset $B$ of $V,$ that is, there exists a positive nondecreasing function $Q$ and a constant $\rho>0$ such that
    \begin{align*}
    dist_V(S(t)B,\mathcal{E})\leq Q(\|B\|_V)e^{-\rho t}
    \end{align*}
    for any $t\geq 0,$ where $dist_V$ denotes the non-symmetric Hausdorff distance between sets in $V$ and $\|B\|_V$ stands for the size of $B$ in $V.$ Moveover, both $Q$ and $\rho$ can be explicitly calculated.
\end{theorem}
\textbf{Proof.} We know from Theorems \ref{3.3}, \ref{4.2.2}, \ref{4.2.12} that there exists a bounded subset $B_0$ in $V$ and $t_1\geq\tau_2+2>0$ such that the mapping $S=S(t_1):B_0\rightarrow B_0$ enjoys the smoothing property
\begin{align}\label{4.2.14}
\|S(v_0^1,T_0^1,q_0^1)-S(v_0^2,T_0^2,q_0^2)\|_V\leq K\|(v_0^1,T_0^1,q_0^1)-(v_0^2,T_0^2,q_0^2)\|_H
 \end{align}
 for any $(v_0^1,T_0^1,q_0^1)$ and $(v_0^1,T_0^1,q_0^1)\in B_0.$

Since $B_0$ is bounded in $V,$ there exists a point $x_0\in B_0$ and a positive constant $R$ such that $B_0\subset B(x_0,R,H),$ where $B(x_0,R,H)$ denotes a $R$-ball in $H$ centered at $x_0\in H.$ We infer from \eqref{4.2.14} that $B(Sx_0,KR,V)$ can cover the image $SB(x_0,R,H).$ Therefore, it follows from the compactness of $V\subset H$ that there exists a finite number of $\theta R$-ball in $H$ with centers $x_1^i$ for any fixed $\theta\in(0,1).$ Moreover, the minimal number of balls in this covering can be estimated as follows:
\begin{align}\label{4.2.15}
N_{\theta R}(B(Sx_0,KR,V),H)=N_{\theta R}(B(0,KR,V),H)=N_{\frac{\theta}{K}}(B(0,1,V),H)=:N(\theta),
 \end{align}
which implies that there exists a finite number $N(\theta)\geq 2$ of $\theta R$-ball in $H$ centered at the points of $V_1=\{x^i_1:i=1,2,\cdots,N(\theta)\}\subset SB_0$ to cover $SB_0$ and
\begin{align*}
dist_H(SB_0,V_1)\leq\theta R.
 \end{align*}
For any $i\in\{1,2,\cdots,N(\theta)\},$ applying the above procedure to every ball $B(x_1^i,\theta R,H),$ we obtain
there exists a finite number $N(\theta)^2$ of $\theta^2 R$-ball in $H$ centered at the points of $V_2=\{x^i_2:i=1,2,\cdots,N(\theta)^2\}\subset S^2B_0$ to cover $S^2B_0$ and
\begin{align*}
dist_H(S^2B_0,V_2)\leq\theta^2 R.
 \end{align*}
Repeating this procedure, we deduce that there exists a finite number $N(\theta)^k$ of $\theta^k R$-ball in $H$ centered at the points of $V_k=\{x^i_k:i=1,2,\cdots,N(\theta)^k\}\subset S^kB_0$ to cover $S^kB_0$ and
\begin{align}\label{4.2.16}
dist_H(S^kB_0,V_k)\leq\theta^k R.
 \end{align}
Now, we define a sequence of sets $E_1=V_1,$ $E_k=SE_{k-1}\cup V_k$ for any $k\in\mathbb{Z}^+.$ Let
\begin{align*}
E=\bigcup_{k=1}^{\infty}E_k
\end{align*}
and let $\mathcal{E}_0$ be the closure of $E$ in $V.$

In what follows, we verify that $\mathcal{E}_0$ is an exponential attractor for $S$ in $V.$ First of all, the invariance follows immediately from our construction. Thanks to $V_k\subset\mathcal{E}_0$ and $\theta\in(0,1),$ we infer from \eqref{4.2.16} that
\begin{align*}
dist_H(S^kB_0,\mathcal{E}_0)\leq dist_H(S^kB_0,V_k)\leq\theta^k R=Re^{ln\theta k },
 \end{align*}
 which implies that
 \begin{align*}
dist_V(S^kB_0,\mathcal{E}_0)\leq\theta^k KR=KRe^{ln\theta k }.
 \end{align*}

Thanks to $SB_0\subset B_0$ and
 \begin{align*}
\bigcup_{k\geq n}E_k\subset S^nB_0\subset\bigcup_{h\in V_n}B(h,\theta^nKR,V).
 \end{align*}
For any $\epsilon>0,$ there exists some smallest positive integer $n$ such that $\theta^nKR\leq\epsilon.$ Therefore, we obtain
 \begin{align*}
N_\epsilon(E,V)\leq &N_\epsilon(\bigcup_{k\geq n}E_k,V)+N_\epsilon(\bigcup_{k<n}E_k,V)\\
\leq&N(\theta)^n+\sum_{k=1}^{n-1}\sharp E_k\\
\leq&N(\theta)^n+\frac{n-1}{N(\theta)-1}-\frac{N(\theta)^2}{(N(\theta)-1)^2}+\frac{N(\theta)^{n+1}}{(N(\theta)-1)^2}\\
\leq&(2+N(\theta))N(\theta)^n,
 \end{align*}
 which implies that
  \begin{align*}
dim_F(E,H)\leq -\frac{ln N(\theta)}{ln\theta}.
 \end{align*}
 It follows from Lemma \ref{4.2.0} and the continuity of $V\subset H$ that
   \begin{align*}
dim_F(E,V)\leq -\frac{ln N(\theta)}{ln\theta},
 \end{align*}
which implies that
    \begin{align*}
dim_F(\mathcal{E}_0,V)\leq -\frac{ln N(\theta)}{ln\theta},
 \end{align*}
Finally, we define
\begin{align*}
\mathcal{E}=\bigcup_{t\in[t_1,2t_1]}S(t)\mathcal{E}_0,
\end{align*}
It is easily verify that $\mathcal{E}$ is an exponential attractor of problem \eqref{2.4}-\eqref{2.14}.\\
 \qed\hfill
\begin{remark}
Thanks to $\mathcal{A}\subset\mathcal{E},$ we infer from Theorem \ref{4.2.13} that the fractal dimension of the global attractor of problem \eqref{2.4}-\eqref{2.12} established in Theorem \ref{4.1.0} is finite.
\end{remark}

\section*{Acknowledgement}
This work was supported by the National Science Foundation of China Grant (11401459) and the Natural Science
Foundation of Shaanxi Province (2015JM1010).

\bibliographystyle{elsarticle-template-num}
\bibliography{BIB}
\end{document}